\documentclass[aps,prl,groupedaddress,superscriptaddress,longbibliography,
twocolumn,
,amsmath,amssymb]{revtex4-2}
\usepackage{color}
\definecolor{darkblue}{rgb}{0.0, 0.0, 0.55}
\usepackage[colorlinks=true,linkcolor=darkblue,anchorcolor=red,citecolor=darkblue, urlcolor=darkblue]{hyperref}
\usepackage{romannum}
\usepackage{physics}
\usepackage{graphicx}
\usepackage{bm}
\usepackage{subfigure}
\usepackage[capitalize]{cleveref}
\usepackage{epstopdf}

\begin{document}

%\title{Quantum-Hall Type Phase Transition in Disordered Axion Insulators}
%\title{Disorder and Global Phase Diagram of an Axion Insulator State}
%\title{Anderson Transition as a Probe of an Axion Insulator State}
\title{Emergent Energy Dissipation in Quantum Limit}
\author{Hailong Li}
% \thanks{Hailong Li and Chui-Zhen Chen are co-first authors}
\affiliation{International Center for Quantum Materials, School of Physics,
Peking University, Beijing 100871, China}
\author{Hua Jiang}
\email{jianghuaphy@suda.edu.cn}
\affiliation{Institute for Advanced Study, Soochow University, Suzhou 215006, China}
\affiliation{Institute for Nanoelectronic Devices and Quantum Computing, Fudan University, Shanghai 200433, China}
\author{Qing-Feng Sun}
% \email{sunqf@pku.edu.cn}
\affiliation{International Center for Quantum Materials, School of Physics,
Peking University, Beijing 100871, China}
% \affiliation{Beijing Academy of Quantum Information Sciences, Beijing 100193, China}
\affiliation{CAS Center for Excellence in Topological Quantum Computation, University of Chinese Academy of Sciences, Beijing 100190, China}
\affiliation{Hefei National Laboratory, Hefei 230088, China}
\author{X. C. Xie}
\email{xcxie@pku.edu.cn}
\affiliation{International Center for Quantum Materials, School of Physics, Peking University, Beijing 100871, China}
% \affiliation{Beijing Academy of Quantum Information Sciences, Beijing 100193, China}
\affiliation{Institute for Nanoelectronic Devices and Quantum Computing, Fudan University, Shanghai 200433, China}
\affiliation{Hefei National Laboratory, Hefei 230088, China}
\date{\today }

\begin{abstract}
Energy dissipation is of fundamental interest and crucial importance in quantum systems.
% However, whether energy dissipation can emerge inside topological systems remains a question, especially when charge transport is topologically protected and quantized.
However, whether energy dissipation can emerge without backscattering inside topological systems remains a question.
As a hallmark, we propose a microscopic picture that illustrates energy dissipation in the quantum Hall (QH) plateau regime of graphene.
Despite the quantization of Hall, longitudinal, and two-probe resistances (dubbed as the quantum limit), we find that the energy dissipation emerges in the form of Joule heat.
It is demonstrated that the non-equilibrium energy distribution of carriers plays much more essential roles than the resistance on energy dissipation.
Eventually, we suggest probing the phenomenon by measuring local temperature increases in experiments and reconsidering the dissipation typically ignored in realistic topological circuits.
% Eventually, we suggest probing the phenomenon by measuring local temperature increases in experiments.
% Furthermore, it is promising to suppress the dissipation within the existing topological circuits and design dissipation-free topological electronics devices.
% \sout{By analyzing the energy distribution of electrons, it is found that electrons can evolve between equilibrium and non-equilibrium without inducing extra two-probe resistance.
% The relaxation of non-equilibrium electrons results in the dissipation of energy along the QH edge states.
% Eventually, we suggest probing the phenomenon by measuring local temperature increases in experiments and reconsidering the dissipation typically ignored in realistic topological circuits.}
\end{abstract}

\maketitle

{\textit{Introduction.}}---
Topology is a basic concept to classify quantum matter of which the electronic band structure is associated with a topological invariant~\cite{mooreBirthTopologicalInsulators2010,hasanColloquiumTopologicalInsulators2010,qiTopologicalInsulatorsSuperconductors2011,bansilColloquiumTopologicalBand2016,qiTopologicalFieldTheory2008}.
One dramatic consequence of such unique electronic systems is the presence of edge or surface states within the band gap and such gapless excitations can support quantized transport response~\cite{hasanColloquiumTopologicalInsulators2010}. 
% which is the so-called bulk-boundary correspondence
Moreover, the corresponding quantized transport could be robust in the sense that the edge states are topologically protected against local perturbations~\cite{qiTopologicalInsulatorsSuperconductors2011}.
As a symbol, two-dimensional topological insulators with time-reversal symmetry breaking, such as integer quantum Hall (QH)~\cite{klitzingNewMethodHighAccuracy1980,thoulessQuantizedHallConductance1982,vonklitzingQuantizedHallEffect1986} and quantum anomalous Hall (QAH) insulators~\cite{yuQuantizedAnomalousHall2010,changExperimentalObservationQuantum2013,liuQuantumAnomalousHall2016,changColloquiumQuantumAnomalous2023,wengQuantumAnomalousHall2015,dengQuantumAnomalousHall2020,haldaneModelQuantumHall1988}, are characterized by Chern number and essentially possess one-dimensional chiral edge states accompanied by exact quantized Hall resistance and zero longitudinal resistance.
The chiral edge states contributing to the exactly quantized charge transport are regarded as dissipationless within the conductor, along which electric currents flow unidirectionally and backscattering is forbidden by topological protection~\cite{buttikerAbsenceBackscatteringQuantum1988,haldaneModelQuantumHall1988}.
% The exact quantization of Hall measurements is traditionally attributed to dissipationless chiral edge states, along which electric currents flow unidirectionally and backscattering is forbidden by topological protection [].
% Such edge states travel unidirectionally and backscattering is topologically forbidden, which turns out to exhibit quantized Hall conductance~\cite{buttikerAbsenceBackscatteringQuantum1988,haldaneModelQuantumHall1988}.
As such, the topological insulators are promising candidates in the next-generation dissipationless electronic devices~\cite{xiaoChiralChannelNetwork2020,wuBuildingProgrammableIntegrated2021,varnavaControllableQuantumPoint2021}.
\begin{figure}[htbp!]
 \centering
 \vspace{-0pt} %调整图片与上文的垂直距离
 \includegraphics[width=0.9\columnwidth]{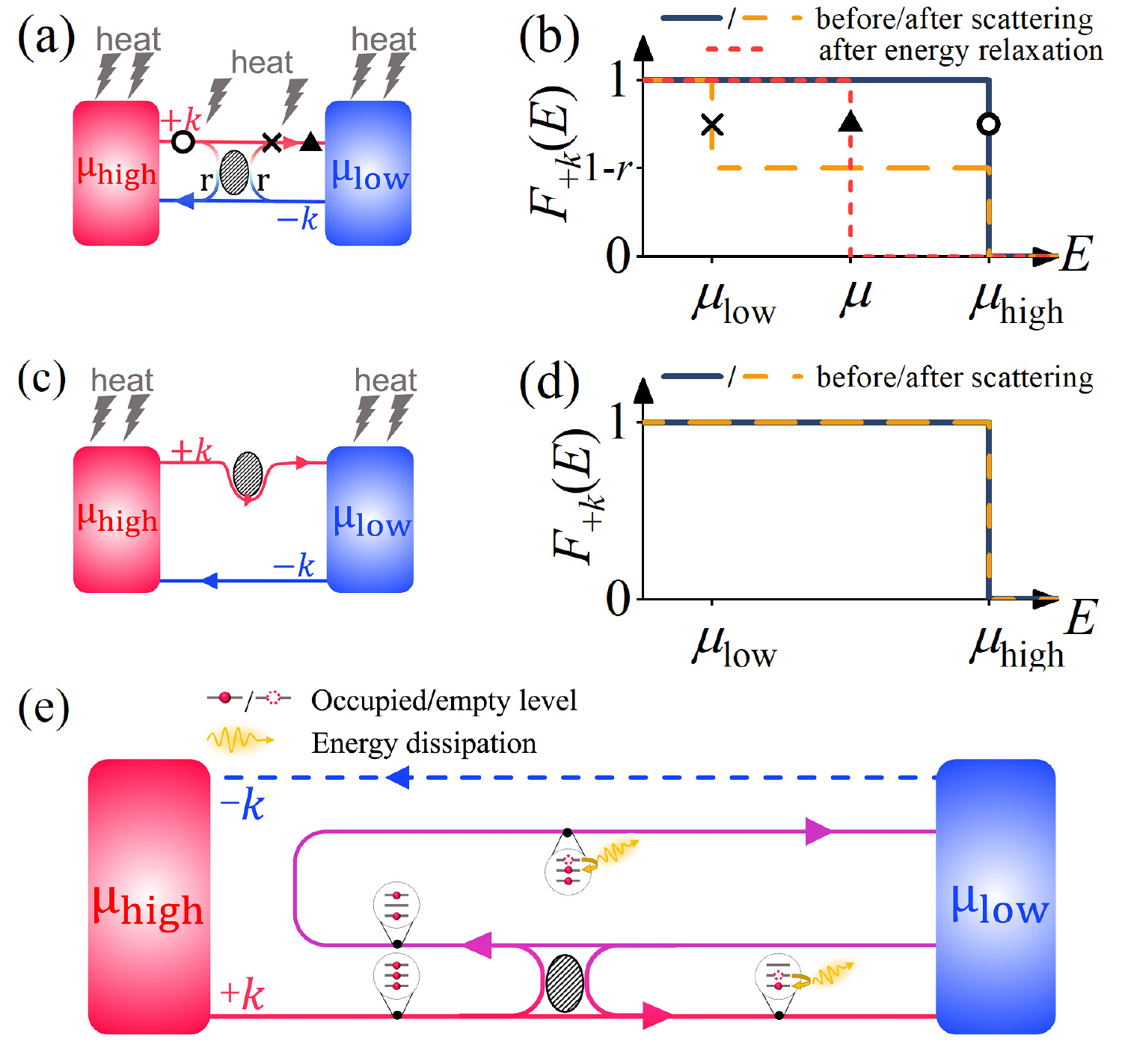}
 \setlength{\abovecaptionskip}{-3pt} %调整图片标题与图距离
 \setlength{\belowcaptionskip}{-15pt} %调整图片标题与下文距离
 \caption{(a) Single-channel conductor with an elastic barrier and (b) energy distribution of right-moving ($+k$) electrons $F_{+k}(E)$ with reflection probability $r$. The symbols in (a) represent electron distribution at various positions, corresponding to the lines in (b).
 In (a), heat is generated in both the contacts and the conductor. The elastic barrier induces a highly non-equilibrium $+k$ state, and heat arises from energy relaxation processes that return this state to a Fermi-distributed state.
 (c) QH/QAH system with an elastic barrier, and (d) energy distribution of $+k$ electrons. The chiral edge channel bypasses the barrier and maintains a Fermi distribution. Consequently, heat generation occurs solely within the contacts.
 In the main text, we find that the conventional interpretation in (a-d) is insufficient to explain the relationship between dissipation and topological protection.
 (e) Schematic illustrating dissipative transport while preserving quantized two-probe resistance. The bubbles qualitatively represent the energy distribution of electrons, revealing how and where non-equilibrium distributions arise and relax.}
 \label{Fig1}
\end{figure}
However, the connection between dissipation and topological protection has not been adequately elucidated in complex topological system~\cite{margueriteImagingWorkDissipation2019,aharon-steinbergLongrangeNontopologicalEdge2021a,fangThermalDissipationQuantum2021,slizovskiyCoolingChiralHeat2017,zhangDissipationResistanceImaging2020}.
Local dissipation within topological devices can result in undesirably high temperatures that can negatively impact device safety, reliability, and performance. 
Thus, we pose the question of whether the topological edge states are immune from heat dissipation in topological quantized charge transport.

A preliminary answer provided by conventional mesoscopic transport theory seems affirmative [see \cref{Fig1}].
For a topologically trivial ballistic conductor with one elastic barrier in~\cref{Fig1}(a), the total resistance ($R=1/(1-r)$) in the units of $h/e^2$ is the sum of the barrier resistance ($R_0=r/(1-r)$) and the intrinsic contact resistance ($R_{\mathrm{c}}=1$), where spin degeneracy is neglected~\cite{landauerSpatialVariationCurrents1957a,landauerElectricalResistanceDisordered1970,imryPHYSICSMESOSCOPICSYSTEMS1986,imryIntroductionMesoscopicPhysics2002}.
Considering inelastic processes, nonequilibrium electrons relax [see~\cref{Fig1}(b)] and energy dissipates in both the contacts and the conductor, which satisfies the Joule's law ($Q=I^2R$)~\cite{SM}.
However, for a QH case with $\nu=1$ [see \cref{Fig1}(c)], the right-moving ($+k$) state will bypass the barrier, leading to a barrier resistance of $R_0=0$~\cite{buttikerAbsenceBackscatteringQuantum1988}.
% always stays in a thermal equilibrium state due to the topological protection~\cite{buttikerAbsenceBackscatteringQuantum1988}. 
Here still exists a minimum contact resistance of $h/e^2$ which is general and inevitable~\cite{imryPHYSICSMESOSCOPICSYSTEMS1986}.
Meanwhile, the topological system enters the quantum limit.
It seems that in the quantum limit, energy dissipation only generates within the contacts rather than inside the conductor.
% , which explicitly captures the features of no dissipation inside the conductor.

However, a recent experiment investigating the QH effect reported unexpected heat dissipation at graphene boundaries which challenges the above statement~\cite{margueriteImagingWorkDissipation2019}. 
% Several theoretical studies proposed some microscopic mechanisms explaining the heat dissipation with quantized Hall resistance~\cite{slizovskiyCoolingChiralHeat2017,zhangDissipationResistanceImaging2020,fangThermalDissipationQuantum2021}. 
Although several theoretical studies proposed some microscopic explanations~\cite{slizovskiyCoolingChiralHeat2017,zhangDissipationResistanceImaging2020,fangThermalDissipationQuantum2021}, they at most demonstrate dissipation can occur under certain backscattering, while the backscattering can keep the Hall conductance quantized but cause additional two-probe resistance.
% However, the systems in these studies do not enter the quantum limit due to their compromise of extra contact resistance from backscattering~\cite{fangThermalDissipationQuantum2021}, or additional heat injection~\cite{slizovskiyCoolingChiralHeat2017,zhangDissipationResistanceImaging2020}.
% One aim of exploring the QH phase is to utilize its property that chiral edge states eliminate local resistance and reach the quantum limit. 
One advantage of chiral edge states is topological protection against backscattering, which in principle leads to a minimum and quantized contact resistance. Such a minimum resistance is always inevitable~\cite{imryPHYSICSMESOSCOPICSYSTEMS1986}.
% Thus, whether energy dissipation can occur in such a situation is still an unanswered question.
Thus, whether energy dissipation can occur in such a situation is not only unresolved, but also more fundamental and important.

\begin{figure}[tp]
 \centering
 \includegraphics[width=\columnwidth]{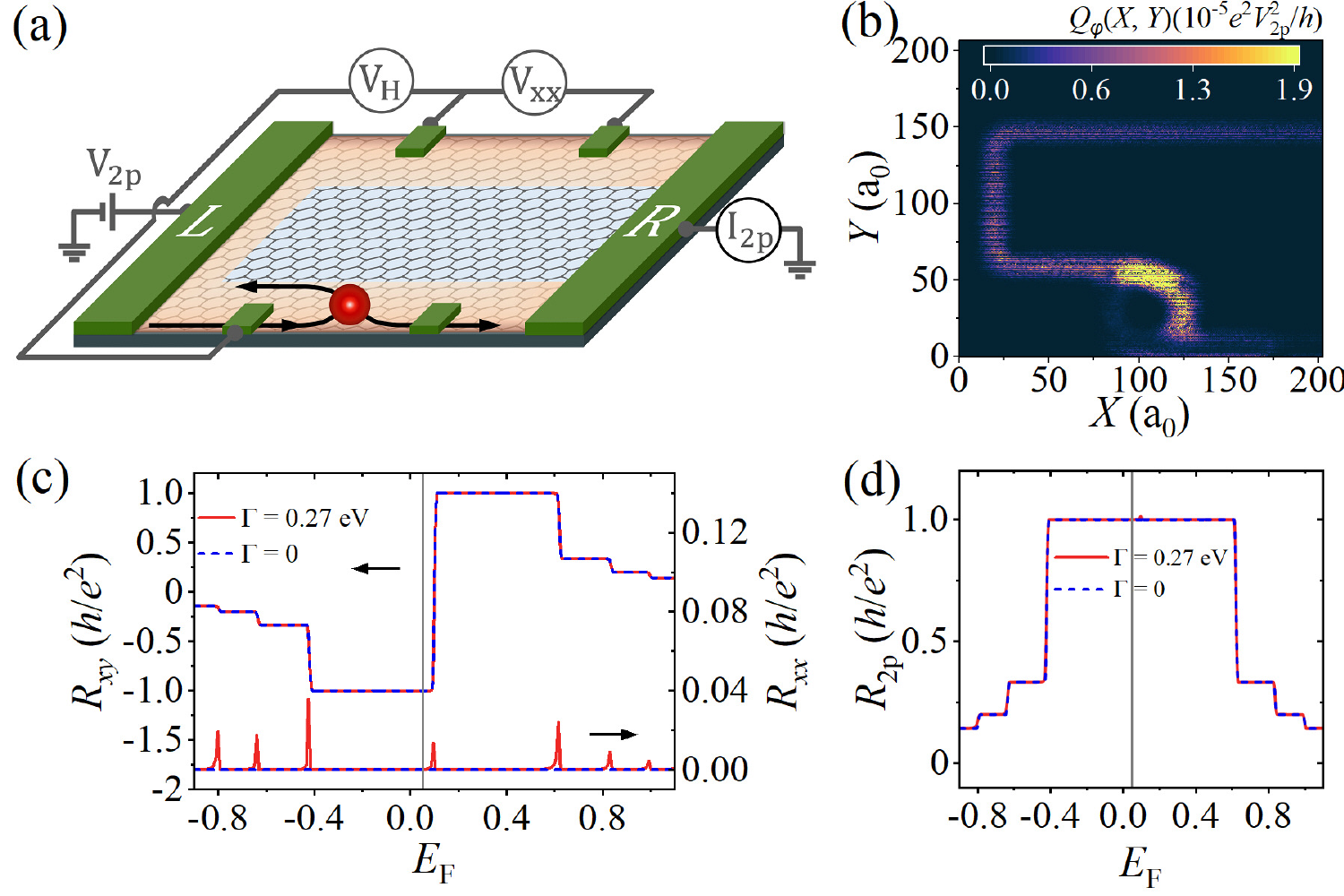}
 \caption{(a). Schematic plot of a six-terminal Hall device in graphene. 
 An edge electrostatic potential $\mathcal{E}^{\mathrm{edge}}_{n}$ is applied in the orange region.
 The inner and outer channels are situated near the inner and outer boundaries of the orange region, respectively.
 An elastic barrier $\mathcal{V}_n$, represented by the red disc near the bottom edge, scatters the edge channels.
 % The chemical potential of contact $\mathrm{L}(\mathrm{R})$ is represented by $\mu_{\mathrm{L}}(\mu_{\mathrm{R}})$.
 A voltage difference $V_{\mathrm{2p}}$ is applied between the left and right contacts and the side contacts are used for voltage measurement.
 (b) The local energy dissipation $Q_{\varphi}(X,Y)$ in the quantum Hall plateau with $R_{xy}=-h/e^2$ and $X, Y$ denote the real space coordinates. Here, the Fermi energy $E_\mathrm{F}=0.05~\mathrm{eV}$ and the external magnetic field $\Phi_{\mathrm{B}}=0.004$.
 (c) Hall resistance $R_{xy}(=V_\mathrm{H}/I_\mathrm{2p})$ and longitudinal resistance $R_{xx}(=V_{xx}/I_\mathrm{2p})$ as a function of $E_\mathrm{F}$ under different dissipation strength $\Gamma$. 
 (d) Two-probe resistance $R_\mathrm{2p}(=V_\mathrm{2p}/I_\mathrm{2p})$ as a function of $E_\mathrm{F}$ under different $\Gamma$.
 The B{$\mathrm{\ddot{u}}$}ttiker probes are randomly distributed all over the graphene sample and the numerical simulation is averaged over 1280 B{$\mathrm{\ddot{u}}$}ttiker-probe configurations.
 }
 \label{Fig2}
\end{figure}

In this work, we propose a microscopic picture in topological systems that illustrates energy dissipation without undermining the quantum limit [see \cref{Fig1}(e)].
Specifically, we investigate the transport properties of graphene in the QH plateau regime in which the Hall, the longitudinal, and the contact/two-probe resistances are all quantized [see \cref{Fig2}(c) and 2(d)].
However, in such a typical scenario, the energy dissipation unexpectedly appears along the propagation of QH edge states, as shown by spatially resolved energy dissipation in \cref{Fig2}(b).
% Furthermore, we elucidate the underlying physics by analyzing the location-dependent distribution of electrons along the path of the QH edge channels which captures where and how the energy dissipation takes place.
Furthermore, we elucidate where and how the energy dissipation takes place by analyzing the energy distribution of electrons along the QH edge channels [see the bubbles in \cref{Fig1}(e)].
To be specific, in the presence of a scattering barrier, electrons are scattered between the channels induced by edge reconstruction.
% Such scattering does not induce any excess two-probe resistance and maintains quantization of the Hall measurement.
Such scattering does not induce any extra two-probe resistance but transforms electrons into non-equilibrium which is necessary for energy dissipation.
% Instead of raising the two-probe resistance or undermining the quantized Hall measurement by backscattering, such scattering in our picture keeps them all quantized.
% Meanwhile, the electron distribution evolves to a highly non-equilibrium distribution which is necessary for heat dissipation.
Through energy relaxation processes, the non-equilibrium electrons dissipate their energy in the form of heat.
% Such thermalizing processes are accompanied by the dissipation of electron energy in the form of Joule heat, which takes place over a long distance.
Ultimately, we discuss the experimental realization and the energy dissipation in existing topological circuits.

{\textit{Emergent dissipation in effective model}}---
We consider a six-terminal Hall measurement device based on graphene [see \cref{Fig2}(a)], of which the spinless tight-binding Hamiltonian reads~\cite{haldaneModelQuantumHall1988,fangThermalDissipationQuantum2021,lewenkopfRecursiveGreenFunction2013,longDisorderInducedEnhancementTransport2008,shengQuantumHallEffect2006,chenDephasingEffectTransport2011},
% \begin{equation} 
% \mathcal{H}=\sum_\mathbf{n}\epsilon_{\mathbf{n}}\hat{c}_{\mathbf{n}}^{\dagger}\hat{c}_{\mathbf{n}}-\sum_{\expval{\mathbf{m},\mathbf{n}}}t {e}^{-\mathrm{i}\varphi_{\mathbf{m}\mathbf{n}}}\hat{c}_{\mathbf{m}}^{\dagger}\hat{c}_{\mathbf{n}}+\hat{\Sigma}_\varphi
% \end{equation}
\begin{equation}
 \mathcal{H}=\sum_n(\mathcal{V}_n+\mathcal{E}^{\mathrm{edge}}_{n})\ \hat{c}_{n}^{\dagger}\hat{c}_{n}-\sum_{\expval{m,n}}t {e}^{-\mathrm{i}\phi_{mn}}\ \hat{c}_{m}^{\dagger}\hat{c}_{n}+\hat{\Sigma}_\varphi
\end{equation}
where $\expval{m,n}$ means the nearest pairs of lattice sites and $t$ is the hopping parameter.
An external magnetic field $B_z$ is included, which is manifested by the Peierls phase $\phi_{mn}$ on the hopping parameter $t$~\cite{wannierDynamicsBandElectrons1962,hofstadterEnergyLevelsWave1976}.
For convenience, the magnetic field is expressed in terms of a dimensionless quantity $\Phi_\mathrm{B}=B_za_0^2/\Phi_0$, with $\Phi_0=h/\abs{e}$ being the magnetic flux quantum and $a_0$ being the lattice constant of graphene~\footnote{Here, $\phi_{mn}=-\int_m^n\Vec{A}\cdot d\Vec{l}/\Phi_0$, where $\Phi_0=h/\abs{e}$ is the magnetic flux quantum and $e<0$ is the electric charge. The vector potential $\Vec{A}$ is chosen as $(-B_zy,0,0)$. Therefore, $\phi_{mn}=-2\pi\Phi_B(x_m-x_n)(y_m+y_n)/2$ with $x_m(x_n)$ and $y_m(y_n)$ denoting the coordinates of site $m(n)$}.

To simulate the elastic scattering process in real devices, we introduce a long-range scattering barrier $\mathcal{V}_n$ near the bottom edge of graphene~\footnote{Here, $\mathcal{V}_n =\mathcal{V}_0e^{-\abs{\mathbf{R}_n-\mathbf{R}_0}/\Lambda}$, where $\mathbf{R}_n=(x_n,~y_n)$ labels the coordinate of lattice site $n$, $\mathbf{R}_0$ denotes the potential center and $\mathcal{V}_0$ describes the scattering strength. In calculations, $\mathbf{R}_0\approx(105a_0, 29.7a_0)$, $\mathcal{V}_0=-0.35\mathrm{eV}$ and $\Lambda=10a_0$}. 
Besides, the edge reconstruction is ubiquitous along the graphene edges induced by charge accumulation~\cite{margueriteImagingWorkDissipation2019,akihoCounterflowingEdgeCurrent2019,chaeEnhancedCarrierTransport2012,silvestrovChargeAccumulationBoundaries2008,cuiUnconventionalCorrelationQuantum2016,panchalVisualisationEdgeEffects2014}.
% some previous findings have suggested the presence of edge reconstruction induced by charge accumulation along the graphene edges~\cite{margueriteImagingWorkDissipation2019,akihoCounterflowingEdgeCurrent2019,chaeEnhancedCarrierTransport2012,silvestrovChargeAccumulationBoundaries2008,cuiUnconventionalCorrelationQuantum2016,panchalVisualisationEdgeEffects2014}.
To incorporate its effect, an electrostatic potential $\mathcal{E}^{\mathrm{edge}}_{n}$ is added in the vicinity of the graphene boundary [see the orange region in \cref{Fig2}(a)].
This approach has demonstrated successful implementation in numerical simulations of giant nonlocality in graphene~\cite{aharon-steinbergLongrangeNontopologicalEdge2021a}.
Energy relaxation processes, such as electron-phonon interaction~\cite{altshullerSuppressionLocalizationEffects1981,dattaElectronicTransportMesoscopic1995} and electron-electron interaction~\cite{altimirasTuningEnergyRelaxation2010,venkatachalamLocalThermometryNeutral2012,itohSignaturesNonthermalMetastable2018}, are also widely present.
Under B{$\mathrm{\ddot{u}}$}ttiker-probe scheme, the complex energy relaxation processes are effectively reduced to a single-particle picture with couplings between electrons and environmental reservoirs~\cite{buttikerCoherentSequentialTunneling1988,golizadeh-mojaradNonequilibriumGreenFunction2007}.
Thus, these processes are involved by attaching B{$\mathrm{\ddot{u}}$}ttiker probes to lattice sites, which manifests itself as an appropriate self-energy functions $\hat{\Sigma}_\varphi=-\mathrm{i}\Gamma/2\sum_n\hat{c}_{n}^{\dagger}\hat{c}_{n}$~\cite{cattenaGeneralizedMultiterminalDecoherent2014,damatoConductanceDisorderedLinear1990}.
We use two parameters to characterize $\hat{\Sigma}_\varphi$, including its strength $\Gamma$ and distribution density $n_\varphi$~\footnote{Here, $n_\varphi$ is defined as the ratio of lattice sites attached with B{$\mathrm{\ddot{u}}$}ttiker probes to the total number of lattice sites.}.
The model parameters for numerical calculations are specified as $t=2.7~\mathrm{eV}$, $\mathcal{E}^{\mathrm{edge}}_n=0.1~\mathrm{eV}$, $\Phi_\mathrm{B}=0.004$, $\Gamma=0.27~\mathrm{eV}$ and $n_{\varphi}=0.005$, unless otherwise specified.

Transport of charge and energy is analyzed via non-equilibrium Green's function method and multi-terminal Landauer-B{$\mathrm{\ddot{u}}$}ttiker formalism, which is expressed under low temperature and low bias~\cite{SM,pourfathNonEquilibriumGreenFunction2014,camsariNonEquilibriumGreenFunction2021}
% Via nonequilibrium Green's function (NEGF) method and multi-terminal Landauer-B{$\mathrm{\ddot{u}}$}ttiker formalism, the electric current $J_{p}$ from terminal $p$ to the conductor and the energy current $Q_{p}$ transferred to terminal $p$ can be both expressed in the linear response regime, i.e., low temperature and low bias~\cite{pourfathNonEquilibriumGreenFunction2014,camsariNonEquilibriumGreenFunction2021,lewenkopfRecursiveGreenFunction2013,cattenaGeneralizedMultiterminalDecoherent2014},
\begin{equation}
 \begin{gathered}
 J_{p}=\frac{e^2}{h}\sum_{q}\mathbf{T}_{pq}(V_p-V_q)\\
 Q_{p}=\frac{e^2}{2h}\sum_{q}\mathbf{T}_{pq}(V_p-V_q)^2-\frac{\pi^2k_\mathrm{B}^2}{6h}\sum_{q}\mathbf{T}_{pq}(T_p^2-T_q^2)
 \end{gathered}\label{eq:current}
\end{equation}
where $J_{p}$ refers to electric current from terminal $p$ to the conductor and $Q_{p}$ refers to energy current transferred to terminal $p$. $p(q)$ are subject to all terminals including real contacts and B{$\mathrm{\ddot{u}}$}ttiker probes~\footnote{For B{$\mathrm{\ddot{u}}$}ttiker probes, electrons exchange their energy under the conservation of particles which means $J_p=0$ there.}. $\mathbf{T}_{pq}$ represents the total transmission from terminal $q$ to terminal $p$. $V_{p(q)}$ and $T_{p(q)}$ refer to the voltage and temperature of terminal ${p(q)}$, respectively.

Thus, based on the six-terminal device and \cref{eq:current}, we numerically calculate the Hall resistance $R_{xy}$ and the longitudinal resistance $R_{xx}$, which shows the system exhibits a perfect QH effect [see \cref{Fig2}(c)].
% Moreover, the calculated two-probe resistance $R_\mathrm{2p}$ shown in \cref{Fig2}(d) also gives a quantized value, $h/e^2$.
Without loss of generality, we assume the dissipated energy is totally transferred to the environmental reservoir via B{$\mathrm{\ddot{u}}$}ttiker probes rather than raises the local temperature, i.e., $T_p=T_0$ for all terminals with the background temperature $T_0=0$. Under such boundary conditions, the dissipated energy appears around the boundary of graphene [see \cref{Fig2}(b)].
To reveal how energy dissipation affects the two-probe resistance $R_\mathrm{2p}$, we choose perfect contacts which electrons can enter without suffering reflections~\cite{dattaElectronicTransportMesoscopic1995,szaferTheoryQuantumConduction1989}.
Here, $R_\mathrm{2p}$ still reaches a minimum value $h/e^2$ in the presence of heat dissipation [see~\cref{Fig2}(d)], which indicates the dissipation inside the conductor is compatible with the quantum limit.
% ~\cite{margueriteImagingWorkDissipation2019,fangThermalDissipationQuantum2021}.
% extra $R_\mathrm{2p}$ is unnecessary to dissipation~\cite{margueriteImagingWorkDissipation2019,fangThermalDissipationQuantum2021}. 
\begin{figure}[b]
 \centering
 \includegraphics[width=\columnwidth]{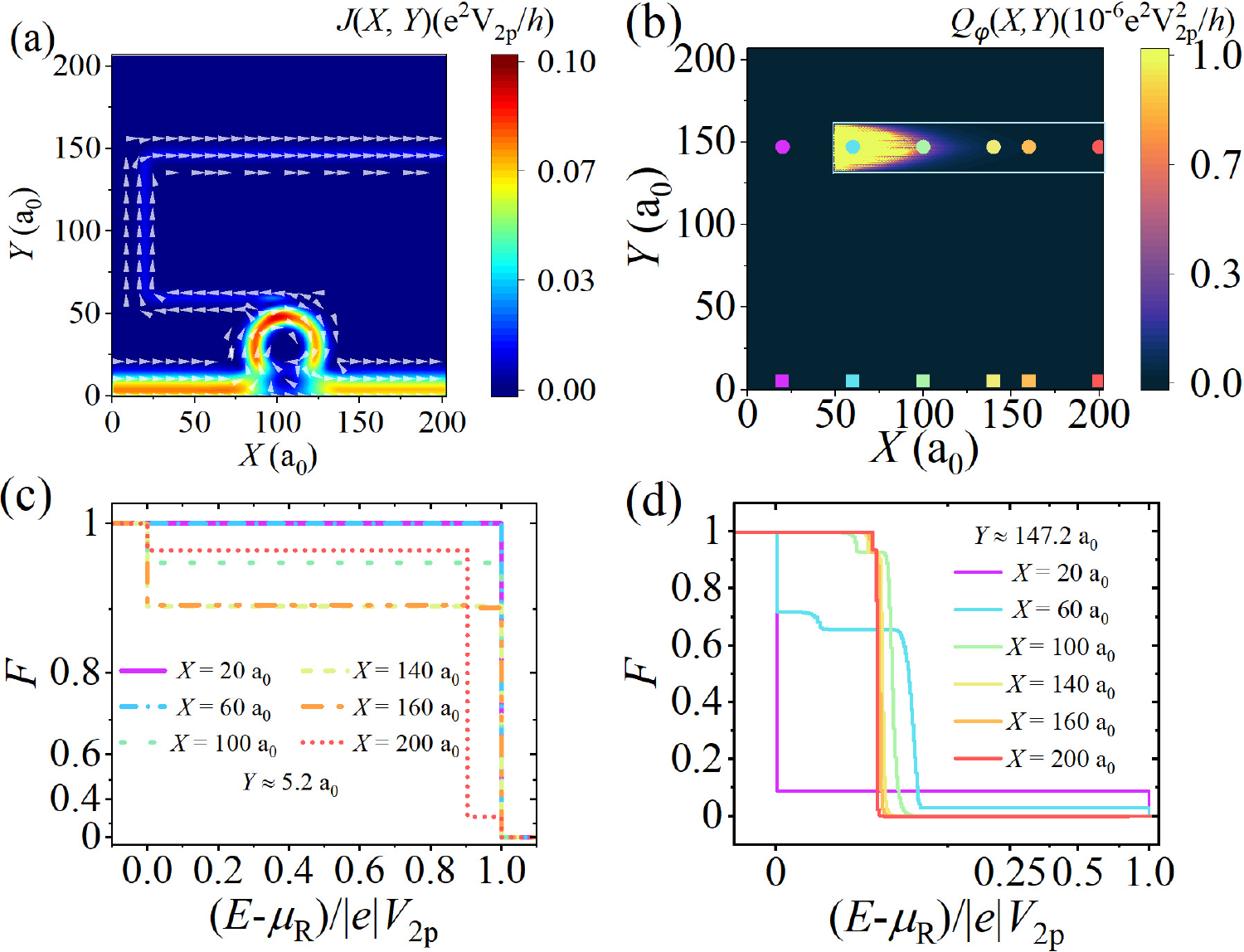}
 \caption{(a) The local electric current $J(X,Y)$ and (b) the local energy dissipation $Q_{\varphi}(X,Y)$ in the quantum Hall plateau regime with $R_{xy}=-h/e^2$.
 Here, $E_\mathrm{F}=0.05~\mathrm{eV}$ and $\Phi_{\mathrm{B}}=0.004$.
 % Here, the Fermi energy $E_\mathrm{F}=0.05$ and the external magnetic field $\Phi_{\mathrm{B}}=0.004$.
 % in the quantum Hall plateau with $R_{xy}=-h/e^2$. 
 In (b), the B{$\mathrm{\ddot{u}}$}ttiker probes are applied in the rectangular region and the strength $\Gamma=0.054~\mathrm{eV}$.
 % In (a) and (b), 
 The energy distribution of electrons $F$ as a function of normalized energy $(E-\mu_\mathrm{R})/\abs{e}V_\mathrm{2p}$ (0 for the right contact and 1 for the left contact ) in (c) and (d) refer to the sites marked by colored squares and dots in (b), respectively.
 }
 \label{Fig3}
\end{figure}

{\textit{Mechanism of energy dissipation under quantum limit}}---
% Non-equilibrium distribution and relaxation of electrons
Initially, we figure out the trajectory of electrons that carry charge and energy in~\cref{Fig2}(b).
The local electric current $J(X,~Y)$~\cite{SM,jauhoTimedependentTransportInteracting1994,nakanishiQuantumLoopCurrent2001,jiangNumericalStudyTopological2009} in \cref{Fig3}(a) captures the flow of electrons in \cref{Fig2}(b) [see S6 of the Supplemental Material for more details].
% Based on the NEGF method, we numerically calculate the local electric current density $J(X,~Y)$~\cite{jauhoTimedependentTransportInteracting1994,nakanishiQuantumLoopCurrent2001,jiangNumericalStudyTopological2009} [see the device setup in \cref{Fig2}(a)], and the flow of the electrons is captured by the result in \cref{Fig3}(a).
% excludes the extra $R_\mathrm{2p}$ introduced by dissipation~\cite{margueriteImagingWorkDissipation2019,fangThermalDissipationQuantum2021}.
% For convenience, the contacts are also made up of graphene with the same magnetic field as the central conductor, which makes the contacts 'perfect', i.e., electrons can enter them without suffering reflections~\cite{dattaElectronicTransportMesoscopic1995,szaferTheoryQuantumConduction1989}.
% It, of course, brings a minimum contact resistance, $h/e^2$~\cite{imryPHYSICSMESOSCOPICSYSTEMS1986,imryIntroductionMesoscopicPhysics2002}.
Due to the presence of edge potential $\mathcal{E}^{\mathrm{edge}}_{n}$, there exists a pair of QH edge channels as inferred by experiments~\cite{margueriteImagingWorkDissipation2019}.
When a bias is applied ($\mu_\mathrm{L}>\mu_\mathrm{R}$), electrons exit from the left contact, enter the central conductor, and populate the outer edge channel.
The scattering barrier $\mathcal{V}_n$ around the bottom edge induces the mixing between the inner and outer edge channels.
After the barrier, the electric current splits into two parts: one moves rightwards along the outer edge channel and the other propagates along the inner channel, both of which end up in the right contact.
Despite the presence of scattering, the total transmission probability of electrons from left to right is still $1$, which leads to the quantum limit of the two-probe resistance $R_\mathrm{2p}=h/e^2$ in \cref{Fig2}(d).

By tracing the path of the electric current, we consider the spatially resolved energy distribution of electrons, $F(n, E)$, to quantitatively address the issue of where and how heat dissipation occurs.
% To answer the questions quantitatively of where and how the heat dissipation occurs, we consider the spatially resolved energy distribution of electrons, $F(n, E)$, tracing the path of the electric current.
% According to the NEGF method, 
Generally, $F(n, E)$ is defined as~\cite{SM,fangThermalDissipationQuantum2021} 
\begin{equation}
 F(n, E)={\Trace[G^\mathrm{e}_{n,n}]}/{\Trace[A_{n,n}]}
\end{equation}
where $A=\mathrm{i}[G^r-G^a]$ is the spectral function and $G^\mathrm{e}_{n,n}=-\mathrm{i}G^{<}_{n,n}$ describes the electron density (times $2\pi$) per unit energy~\cite{dattaElectronicTransportMesoscopic1995}.
To concisely show the underlying physics, B{$\mathrm{\ddot{u}}$}ttiker probes are only placed in the rectangle region [see \cref{Fig3}(b)]~\footnote{Here, the choice of B{$\mathrm{\ddot{u}}$}ttiker probes configuration does not change the Hall result as in~\cref{Fig2} but redistributes the energy dissipation. The configuration in~\cref{Fig3}(b) makes it easier to clarify the thermodynamic evolution of the electron distribution before and after the scattering barrier.}.
% Therefore, heat dissipation is expected to appear only in the same rectangle region.

\begin{figure}[t]
 \centering
 \includegraphics[width=1\columnwidth]{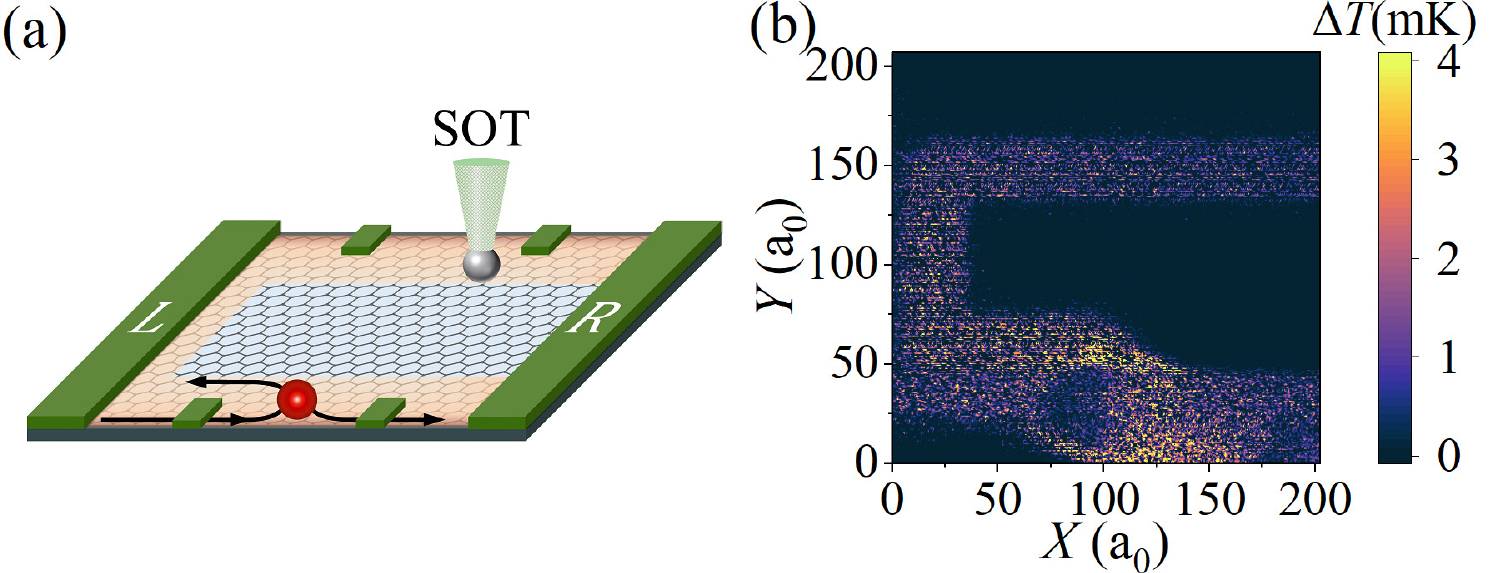}
 \caption{(a) Schematic diagram of the temperature measurement with a superconducting quantum interference device on tip.
 (b) The local temperature increase $\Delta T$ as a function of the location $(X, Y)$. 
 Here, $T_0$, the background temperature, is set to be $4.2~\mathrm{K}$.
 % $0.1\abs{e}V_\mathrm{2p}/k_\mathrm{B}$ where $V_\mathrm{2p}$ is the two-probe voltage in (a) and $k_\mathrm{B}$ is the Boltzmann constant. 
 The B{$\mathrm{\ddot{u}}$}ttiker probes are randomly distributed all over the graphene sample and the numerical simulation is averaged over 1280 B{$\mathrm{\ddot{u}}$}ttiker-probe configurations. }
 \label{Fig4}
\end{figure}

First, we trace the outer channel at the bottom edge, as marked by colored squares in \cref{Fig3}(b), and plot the distribution in \cref{Fig3}(c).
To the left of the barrier, the right-moving channel is fully populated by electrons from the left contact since the contact is ``perfect"~\cite{dattaElectronicTransportMesoscopic1995,szaferTheoryQuantumConduction1989}.
These electrons stay in equilibrium with the left contact, i.e., a zero-temperature Fermi distribution with chemical potential $\mu_\mathrm{L}$~\cite{sivanMultichannelLandauerFormula1986}.
% of which the distribution function reads $F(E)=\theta(\mu_\mathrm{L}-E)$ at zero temperature.
This semiclassical analysis is consistent with the numerical result in \cref{Fig3}(c) at $x\lesssim 60 a_0$ which is exactly a step function $F(E)=\theta(\mu_\mathrm{L}-E)$.
Under the action of the barrier, electrons from the outer channel with a chemical potential $\mu_\mathrm{L}$ are scattered with a probability into the inner channel where electrons are initially in equilibrium with a chemical potential $\mu_\mathrm{R}(<\mu_\mathrm{L})$, 
% with the inner channel with chemical potential $\mu_\mathrm{R}$, and
Thus, the scattered right-moving state is partially occupied within the energy interval ($\mu_\mathrm{R}$,~$\mu_\mathrm{L}$), which means the distribution function is no longer a step function.
Such a non-equilibrium distribution is numerically confirmed in \cref{Fig3}(c) at $x\gtrsim 100 a_0$.
% This situation is also demonstrated by highly non-equilibrium distribution functions in \cref{Fig3}(c) at $x\gtrsim 100 a_0$ where $F(E)'s$ are not step functions any longer.
The above elastic scattering induced by the barrier generates non-equilibrium electrons which are necessary to heat generation~\footnote{To the right of the barrier, electrons in the outer channel are out of equilibrium. With globally distributed B{$\mathrm{\ddot{u}}$}ttiker probes, the electrons relax and the energy dissipates in the form of heat. Contrarily, in~\cref{Fig3}(b), without B{$\mathrm{\ddot{u}}$}ttiker probe at the bottom edge, electrons can not relax and no heat dissipation is observed.}. We also investigate a case without the elastic barrier $\mathcal{V}_n$ in the Supplemental Material~\cite{SM}, in which the absence of heat dissipation confirms that the energy dissipation is indeed sourced from the edge states.

Next, we consider the inner channel marked by colored dots in \cref{Fig3}(b) which passes through the rectangle region attached with B{$\mathrm{\ddot{u}}$}ttiker probes.
Similarly, electrons in the inner channel after the barrier are also out of equilibrium.
In the rectangle region, non-equilibrium electrons exchange their energy instead of particles with environmental reservoirs. 
Through this irreversible process, equilibrium reestablishes and the corresponding distribution of electrons is expected to gradually restore to a Fermi function with a new chemical potential.
Meanwhile, the energy dissipates into the environmental reservoirs in the form of heat.
Such a relaxation process is verified by distribution functions in \cref{Fig3}(d) at $x\gtrsim 20 a_0$ where $F(E)$ gradually evolves back to a step function at $x\approx 200 a_0$.
The dissipated energy is also presented in~\cref{Fig3}(b).
In brief, we demonstrate that energy dissipation is compatible with the quantum limit.

{\textit{Experimental routines and discussions}}---
On the experiment side, the energy dissipation inevitably leads to an increase in local temperature, due to the finite thermal conductivity between the device and the environment.
Thus, measuring the variation of local temperature $\Delta T$ is a feasible scheme to verify our picture.
Recently, a scanning nano-thermometer has been reported to probe local temperature variations with high sensitivity at low temperature, which utilizes a superconducting quantum interference device on a tip, or namely, SOT~\cite{vasyukovScanningSuperconductingQuantum2013,halbertalNanoscaleThermalImaging2016}.
As for experimental routines, we consider a concrete model in \cref{Fig4}(a) which depicts a thermometer SOT placed above the six-terminal graphene device.
% With the result of heat dissipation and SOT, here we discuss a connection between our analysis and possible experimental realization.

% In a realistic situation, the thermal conductivity between the device and the environment is not always good, which indicates the inevitable increase in local temperature.
To simulate the variation of local temperature, we randomly assign some of the B{$\mathrm{\ddot{u}}$}ttiker probes with poor thermal conductivity, i.e., $Q_{p}=0$ and $T_p\neq T_0$ for certain $p$.
Utilizing the boundary conditions and \cref{eq:current}, the local temperature variation $\Delta T$ is illustrated in \cref{Fig4}(b), while the device maintains quantized resistance of $h/e^2$ in the quantum limit.
% quantized charge transport.
% Such a map of local temperature increases verifies our picture that non-equilibrium electrons relax into equilibrium along the path in Fig2a.
% Its shows a good agreement with our picture.
% In particular, the heated region locates around the edge reconstruction region which is basically consistent with the spatial distribution of dissipated energy in \cref{Fig2}(b).
Notably, the region where temperature increases in \cref{Fig4}(b) basically corresponds to the region where energy dissipates, as shown in \cref{Fig2}(b).
% which verifies our picture that non-equilibrium electrons relax into equilibrium along the path in Fig2a.
The commonly accepted view that heat generation in the quantum limit occurs only at the contacts and not within the conductor is challenged by the rise in local temperature inside the conductor.
% and quantized charge transport predicted by our picture verify that energy dissipation is compatible with quantized charge transport, especially, at no cost of extra two-probe resistance.

% Within limits of numerical performance, the edge width is set to be about $60a_0$ ($14.82~\mathrm{nm}$).
Quantitatively, the enhancement of local temperature $\Delta T$ in \cref{Fig4}(b) is up to three orders of magnitude less than the background temperature $T_0=4.2~\mathrm{K}$~\footnote{In numerical calculation, $T_0=0.1\abs{e}V_{\mathrm{2p}}/k_\mathrm{B}$ and the two-probe voltage difference $V_{\mathrm{2p}}=3.6~\mathrm{mV}$.}.
Remarkably, $\Delta T$ can be captured by the micro-kelvin sensitivity of the nano-thermometer SOT~\cite{margueriteImagingWorkDissipation2019,halbertalNanoscaleThermalImaging2016}.
Besides, the width of edge reconstruction in graphene experiments is about several hundreds of nanometers, for instance, about $260~\mathrm{nm}$ in the QH transport~\cite{margueriteImagingWorkDissipation2019} and $200~\mathrm{nm}$-width edge charge accumulation region in the nonlocal transport~\cite{aharon-steinbergLongrangeNontopologicalEdge2021a}.
Experimentally, it is feasible to reversibly write high-resolution doping patterns and artificially tune the edge potential~\cite{shireversiblewritinghighmobility2020}.
Under an external magnetic field, $B\sim 1~\mathrm{Tesla}$, the width scale of QH edge states is in the magnitude order of the magnetic length $l_\mathrm{B}=\sqrt{{\hbar}/{\abs{e}B}}\sim 60~\mathrm{nm}$.
Thus, it is feasible to fabricate a device as in~\cref{Fig4}(a) and probe the emergent energy dissipation by quantifying the local temperature increments as anticipated by our theoretical framework.
% It is, thus, feasible to accommodate the counter-propagating pair of QH edge states in graphene.
% Thus, we propose probing the energy dissipation by measuring local temperature variations.

On the practical side, our analysis provides further insight into topological circuit designs.
% Recently, several groups have reported the realization of a QAH junction with different Chern numbers in magnetic topological multilayer heterostructures~\cite{zhaoCreationChiralInterface2023} and layered topological magnet $\mathrm{MnBi_2Te_4}$~\cite{ovchinnikovTopologicalCurrentDivider2022}.
Recently, several groups have reported the realization of a QAH junction with different Chern numbers~\cite{zhaoCreationChiralInterface2023,ovchinnikovTopologicalCurrentDivider2022}.
Such Chern insulator junctions can function as a chiral edge-current divider, and even facilitate the development of topological circuits~\cite{wuBuildingProgrammableIntegrated2021,varnavaControllableQuantumPoint2021,xiaoChiralChannelNetwork2020}.
As an example, we perform a calculation on one of the devices in the Supplemental Material~\cite{SM}.
While the quantization of charge transport may not be compromised by the scattering among multiple chiral channels, as reported in recent studies~\cite{zhaoCreationChiralInterface2023,ovchinnikovTopologicalCurrentDivider2022}, energy dissipation can still appear according to our picture~\cite{SM}.
In other words, for many topological devices in the quantum limit~\cite{yasudaquantizedchiraledge2017,rosenchiraltransportmagnetic2017,varnavaControllableQuantumPoint2021,wuBuildingProgrammableIntegrated2021,abaninquantizedtransportgraphene2007,jielectronicmachzehnder2003}, energy dissipation can still emerge inside the conductors, which is beyond Ref.~\cite{fangThermalDissipationQuantum2021}.
Local dissipation inside topological devices may cause excessive temperature which is harmful to their safety, reliability, and performance.
% It is also important to note that the generation of heat dissipation will occur in certain regions along the edge channels, according to our physical picture.
As such, it still requires careful attention to energy dissipation in topological quantum systems and topological quantum computation.

{\textit{Conclusion.}}---
% We propose a microscopic picture elucidating that energy dissipation can emerge inside topological quantum systems. 
To summarize, we find that the energy of electrons can dissipate without backscattering, while chiral edge states reduce the resistance to a minimum and quantized value.
In the QH plateau regime of graphene, energy dissipation emerges in the form of heat.
Remarkably, the generation of energy dissipation takes place in the quantum limit, i.e., $R_\mathrm{2p}=h/e^2$, $R_{xy}=-h/e^2$ and $R_{xx}=0$.
The analysis of electron energy distribution reveals that electrons can evolve between equilibrium and non-equilibrium states which causes energy dissipation along the QH edge states without introducing any extra two-probe resistance $R_\mathrm{2p}$.
% Thus, our work demonstrates that energy dissipation is compatible with the quantum limit, which can be probed by measuring local temperature increases, and also provides further comprehension for future topological circuit designs.
Such dissipation can be probed by measuring local temperature increase.
Furthermore, our work demonstrates that energy dissipation without backscattering is especially for the application of topological materials in low-dissipation electronic devices.

{\textit{Acknowledgement.}}---
% sun: 11921005;Xie: XDB28000000;JiangHua: 2019YFA0308403;
We thank Haiwen Liu, Jie Liu, Qing Yan, and Jing-Yun Fang for illuminating discussions.
This work is financially supported by the Innovation Program for Quantum Science and Technology (Grants No. 2021ZD0302400), the National Key R\&D Program of China (Grants No. 2019YFA0308403 and No. 2022YFA1403700), the National Natural Science Foundation of China (Grants No. 11921005, No. 12304052 and No. 12374034), and the Strategic Priority Research Program of the Chinese Academy of Sciences (Grant No. XDB28000000). Hailong Li is also funded by China Postdoctoral Science Foundation (Grant No. BX20220005).
% the National Basic Research Program of China (Grant No. 2019YFA0308403), and the National Natural Science Foundation of China under Grants No. 11822407 and No. 11974256. C.-Z. C. is also funded by the Natural Science Foundation of Jiangsu Province under Grant No. BK20190813.

{\textit{Note added.}---The computer programs associated with this paper are available via the GitHub website} (\url{https://github.com/meplum-li/dissipation_calc.git}).

% \bibliography{reference.bib}


\end{document}

\title{Supplementary Materials for ``Emergent Energy Dissipation in Quantum Limit''}
\author{Hailong Li}
% \thanks{Hailong Li and Chui-Zhen Chen are co-first authors}
\affiliation{International Center for Quantum Materials, School of Physics,
Peking University, Beijing 100871, China}
\author{Hua Jiang}
\email{jianghuaphy@suda.edu.cn}
\affiliation{Institute for Advanced Study, Soochow University, Suzhou 215006, China}
\affiliation{Institute for Nanoelectronic Devices and Quantum Computing, Fudan University, Shanghai 200433, China}
\author{Qing-Feng Sun}
% \email{sunqf@pku.edu.cn}
\affiliation{International Center for Quantum Materials, School of Physics,
Peking University, Beijing 100871, China}
% \affiliation{Beijing Academy of Quantum Information Sciences, Beijing 100193, China}
\affiliation{CAS Center for Excellence in Topological Quantum Computation, University of Chinese Academy of Sciences, Beijing 100190, China}
\affiliation{Hefei National Laboratory, Hefei 230088, China}
\author{X. C. Xie}
\email{xcxie@pku.edu.cn}
\affiliation{International Center for Quantum Materials, School of Physics, Peking University, Beijing 100871, China}
% \affiliation{Beijing Academy of Quantum Information Sciences, Beijing 100193, China}
\affiliation{Institute for Nanoelectronic Devices and Quantum Computing, Fudan University, Shanghai 200433, China}
\affiliation{Hefei National Laboratory, Hefei 230088, China}
\date{\today }

\maketitle
\tableofcontents
\clearpage
\section{Semiclassical analysis of Joule's heat}
\begin{figure}[htbp]
 \centering
 \includegraphics[width=1\columnwidth]{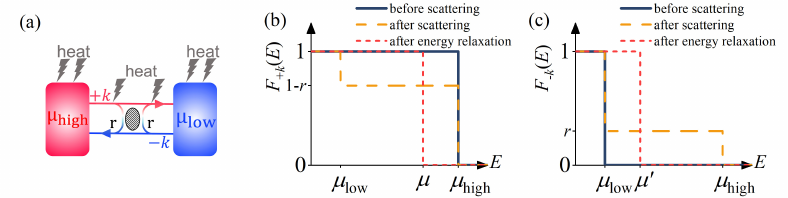}
 \caption{(a) Single-channel ballistic conductor with one elastic barrier. 
 Energy distribution of the right-moving ($+k$) electrons $F_{+k}(E)$ (b) and the left-moving electrons $F_{-k}(E)$ (c) with reflection probability $r$. In (a), the generation of heat occurs in both the contacts and the conductor. The elastic barrier makes the $+k$ and $-k$ states highly non-equilibrium and the heat occurs through energy relaxation processes which evolve the highly non-equilibrium state back into a Fermi-distributed state.}
 \label{fig:SM-1}
\end{figure}
To reveal how energy dissipates in the form of Joule heat, we choose perfect contacts which electrons can enter without suffering reflections~\cite{dattaElectronicTransportMesoscopic1995,szaferTheoryQuantumConduction1989,imryPHYSICSMESOSCOPICSYSTEMS1986,imryIntroductionMesoscopicPhysics2002}.
In the following analysis, we will neglect interference effects and treat the electrons as semiclassical particles.
In~\cref{fig:SM-1}, the $+k$ states to the left of the barrier are occupied only by electrons coming in from the left contact. At zero temperature, the distribution function reads [black line in~\cref{fig:SM-1}(b)]
\begin{equation}
 F_{+k}^{(1)}(E)=f_{\mathrm{left}}(E)=\theta(\mu_{\mathrm{high}}-E)\quad\quad\quad\quad\text{($+k$ before scattering)}\label{eq:f+k1}
\end{equation}
Similarly, the chemical potential of particles occupying $-k$ states to the right of the barrier is equivalent to that of the right contact. Hence, the distribution function of $-k$ is given by [black line in~\cref{fig:SM-1}(c)]\
\begin{equation}
 F_{-k}^{(1)}(E)=f_{\mathrm{right}}(E)=\theta(\mu_{\mathrm{low}}-E)\quad\quad\quad\quad\text{($-k$ before scattering)}\label{eq:f-k1}
\end{equation}
For $+k$ states after scattering but before energy relaxation, the states between $\mu_{\mathrm{low}}$ and $\mu_{\mathrm{high}}$ are partially filled with probability $(1-r)$ (the transmission probability $1-r$), so that [yellow line in~\cref{fig:SM-1}(b)]
\begin{equation}
 F_{+k}^{(2)}(E)=\theta(\mu_{\mathrm{low}}-E)+(1-r)\left[\theta(\mu_{\mathrm{high}}-E)-\theta(\mu_{\mathrm{low}}-E)\right]\quad\quad\text{($+k$ after scattering but before energy relaxation)}\label{eq:f+k2}
\end{equation} 
Similarly, for $-k$ states after scattering but before energy relaxation, the states between $\mu_{\mathrm{low}}$ and $\mu_{\mathrm{high}}$ are partially filled with probability $(r)$, so that [yellow line in~\cref{fig:SM-1}(c)]
\begin{equation}
 F_{-k}^{(2)}(E)=\theta(\mu_{\mathrm{low}}-E)+r\left[\theta(\mu_{\mathrm{high}}-E)-\theta(\mu_{\mathrm{low}}-E)\right]\quad\quad\text{($-k$ after scattering but before energy relaxation)}\label{eq:f-k2}
\end{equation} 
$F_{+k}^{(2)}(E)$ and $F_{-k}^{(2)}(E)$ are highly non-equilibrium distributions. After energy relaxation, they become [red line in~\cref{fig:SM-1}(b)]
\begin{equation}
 F_{+k}^{(3)}(E)=\theta(\mu-E)\quad\quad\quad\quad\text{($+k$ after energy relaxation)}\label{eq:f+k3}
\end{equation}
and [red line in~\cref{fig:SM-1}(c)]
\begin{equation}
 F_{-k}^{(3)}(E)=\theta(\mu'-E)\quad\quad\quad\quad\text{($-k$ after energy relaxation)}\label{eq:f-k3}
\end{equation}
Due to the conservation of particles, i.e., the total number of electrons is invariant, the new $\mu$ and $\mu'$ reads
\begin{align}
 \mu&=\mu_{\mathrm{low}}+(1-r)(\mu_{\mathrm{high}}-\mu_{\mathrm{low}})\label{eq:mu}\\
 \mu'&=\mu_{\mathrm{low}}+r(\mu_{\mathrm{high}}-\mu_{\mathrm{low}})\label{eq:mu'}
\end{align}

The electric current flows from the left contact to the right contact is determined by
\begin{equation}
 \begin{split}
 I&=\frac{e}{h}\int_{-\infty}^{+\infty}T(E)(f_{\mathrm{left}}(E)-f_{\mathrm{right}}(E))dE\\
 &=\frac{e}{h}\int_{-\infty}^{+\infty}(1-r)(\theta(\mu_{\mathrm{high}}-E)-\theta(\mu_{\mathrm{low}}-E))dE\\
 &=-\frac{e^2}{h}(1-r)\Delta V 
 \end{split} 
\end{equation}
Here, $-e\Delta V=\mu_{\mathrm{high}}-\mu_{\mathrm{low}}$ and $e<0$. Thus, the total resistance $R=1/(1-r)$ in the unit of $h/e^2$ which can be written as a series combination of contact resistance ($R_c=1$) and the barrier resistance ($R_0=r/(1-r)$).

Next, we analyze the energy dissipation which consists of four different parts, including relaxation of $+k$ states in the interior of the conductor ($Q_{+k}$), relaxation of $-k$ states in the interior of the conductor ($Q_{-k}$), relaxation at the interfaces near the left ($Q_{\mathrm{L}}$) and right ($Q_{\mathrm{R}}$) contacts. 
\begin{enumerate}
 \item \textbf{Relaxation at the interfaces near the left contact, $Q_{\mathrm{L}}$}

 According to the semiclassical analysis of the distribution functions, $+k$ electrons with $F_{+k}^{(1)}(E)$ flows out of the left contact and $-k$ electrons with $F_{-k}^{(3)}(E)$ flow into the left contact. 
 For instance, electrons with positive momentum $+k$ flowing in the conductor with energy $E$ are injected from the left contact with energy $\mu_{\mathrm{high}}$. Here, the left contact serves as an electron reservoir with chemical potential $\mu_{\mathrm{high}}$. In this process, the loss of energy $\mu_{\mathrm{high}}-E$ is dissipated.
 Thus, from~\cref{eq:f+k1} and~\cref{eq:f-k3} the power of energy dissipation is
 \begin{equation}
 \begin{split}
 Q_{\mathrm{L}}&=\frac{1}{h}\int_{-\infty}^{+\infty}(\mu_{\mathrm{high}}-E)\left(F_{+k}^{(1)}(E)-F_{-k}^{(3)}(E) \right)\mathrm{d}E\\
 &=\frac{1}{h}\int_{\mu'}^{\mu_{\mathrm{high}}}(\mu_{\mathrm{high}}-E)\mathrm{d}E\\
 &=\frac{1}{2h}\left(\mu'-\mu_{\mathrm{high}}\right)^2\quad\quad\text{(from~\cref{eq:mu'})}\\
 &=\frac{e^2}{2h}(1-r)^2(\Delta V)^2
 \end{split}
 \end{equation}
 \item \textbf{Relaxation at the interfaces near the right contact, $Q_{\mathrm{R}}$}
 
 Similarly, $-k$ electrons with $F_{-k}^{(1)}(E)$ flows out of the right contact and $+k$ electrons with $F_{+k}^{(3)}(E)$ flow into the right contact. Thus, from~\cref{eq:f+k3} and~\cref{eq:f-k1}, the power of energy dissipation is
 \begin{equation}
 \begin{split}
 Q_{\mathrm{R}}&=\frac{1}{h}\int_{-\infty}^{+\infty}(E-\mu_{\mathrm{low}})\left(F_{+k}^{(3)}(E)-F_{-k}^{(1)}(E) \right)\mathrm{d}E\\
 &=\frac{1}{h}\int_{\mu_{\mathrm{low}}}^{\mu}(E-\mu_{\mathrm{low}})\mathrm{d}E\\
 &=\frac{1}{2h}\left(\mu-\mu_{\mathrm{low}}\right)^2\quad\quad\text{(from~\cref{eq:mu})}\\
 &=\frac{e^2}{2h}(1-r)^2(\Delta V)^2
 \end{split}
 \end{equation}
 \item \textbf{Relaxation of $+k$ states in the interior of the conductor, $Q_{+k}$}

 It refers to the evolution of the non-equilibrium distribution $F_{+k}^{(2)}(E)$ to the Fermi distribution $F_{+k}^{(3)}(E)$. From~\cref{eq:f+k2} and~\cref{eq:f+k3} the power of energy dissipation is
 \begin{equation}
 \begin{split}
 Q_{+k}&=\frac{1}{h}\int_{-\infty}^{+\infty}(E-\mu)\left(F_{+k}^{(2)}(E)-F_{+k}^{(3)}(E) \right)\mathrm{d}E\\
 &=\frac{1}{h}\int_{\mu_{\mathrm{low}}}^{\mu_{\mathrm{high}}}(E-\mu)(1-r)\mathrm{d}E-\frac{1}{h}\int_{\mu_{\mathrm{low}}}^{\mu}(E-\mu)\mathrm{d}E\\
 &=\frac{1}{2h}(1-r)[\left(\mu_{\mathrm{high}}-\mu\right)^2-\left(\mu_{\mathrm{low}}-\mu\right)^2]+\frac{1}{2h}\left(\mu_{\mathrm{low}}-\mu\right)^2\\
 &=\frac{e^2}{2h}r(1-r)(\Delta V)^2
 \end{split}
 \end{equation}
 \item \textbf{Relaxation of $-k$ states in the interior of the conductor, $Q_{-k}$}

 It refers to the evolution of the non-equilibrium distribution $F_{-k}^{(2)}(E)$ to the Fermi distribution $F_{-k}^{(3)}(E)$. From~\cref{eq:f-k2} and~\cref{eq:f-k3} the power of energy dissipation is
 \begin{equation}
 \begin{split}
 Q_{-k}&=\frac{1}{h}\int_{-\infty}^{+\infty}(E-\mu')\left(F_{-k}^{(2)}(E)-F_{-k}^{(3)}(E) \right)\mathrm{d}E\\
 &=\frac{1}{h}\int_{\mu_{\mathrm{low}}}^{\mu_{\mathrm{high}}}(E-\mu')r\mathrm{d}E-\frac{1}{h}\int_{\mu_{\mathrm{low}}}^{\mu'}(E-\mu')\mathrm{d}E\\
 &=\frac{1}{2h}r[\left(\mu_{\mathrm{high}}-\mu'\right)^2-\left(\mu_{\mathrm{low}}-\mu'\right)^2]+\frac{1}{2h}\left(\mu_{\mathrm{low}}-\mu'\right)^2\\
 &=\frac{e^2}{2h}r(1-r)(\Delta V)^2
 \end{split}
 \end{equation}
\end{enumerate}
Therefore, the total power of energy dissipation is 
 \begin{equation}
 \begin{split}
 Q&=Q_{\mathrm{L}}+Q_{\mathrm{R}}+Q_{+k}+Q_{-k}\\
 &=\frac{e^2}{2h}(1-r)^2(\Delta V)^2+\frac{e^2}{2h}(1-r)^2(\Delta V)^2+\frac{e^2}{2h}r(1-r)(\Delta V)^2+\frac{e^2}{2h}r(1-r)(\Delta V)^2\\
 &=\frac{e^2}{h}(1-r)(\Delta V)^2\\
 &=I^2 R\quad\quad\quad\quad \text{$R=\frac{h}{e^2(1-r)}$ is the total resistance}
 \end{split}
 \end{equation}
In other words, the total power of energy dissipation satisfies Joule's law. Moreover, the power of energy dissipation within the conductor is determined by $Q_{+k}$ and $Q_{-k}$
\begin{equation}
 \begin{split}
 Q_{0}&=Q_{+k}+Q_{-k}\\
 &=\frac{e^2}{2h}r(1-r)(\Delta V)^2+\frac{e^2}{2h}r(1-r)(\Delta V)^2\\
 &=\frac{e^2}{h}r(1-r)(\Delta V)^2\\
 &=(-\frac{e^2}{h}(1-r)\Delta V )^2\times \frac{rh}{(1-r)e^2}\\
 &=I^2 R_{0}\quad\quad\quad\quad \text{$R_{0}=\frac{rh}{(1-r)e^2}$ is the barrier resistance}
 \end{split}\label{eq:q0}
\end{equation}

According to \cref{eq:q0}, as long as $r\neq 0$ (i.e., backscattering exists), $Q_0>0$, and energy dissipation will occur within the conductor.
From the conventional viewpoint, topological systems with chiral edge states can eliminate local resistance (i.e., $r=0$ and $R_0=0$) and leave only quantized contact or terminal resistance. In this way, it is claimed that dissipation-free transport can be realized. 
However, the validity of this assertion remains an unanswered question which also constitutes the primary focus of our research.
\clearpage
\pagebreak
\section{A case in the absence of the elastic barrier $\mathcal{V}_n$}
\begin{figure}[htbp]
 \includegraphics[width=.6\columnwidth]{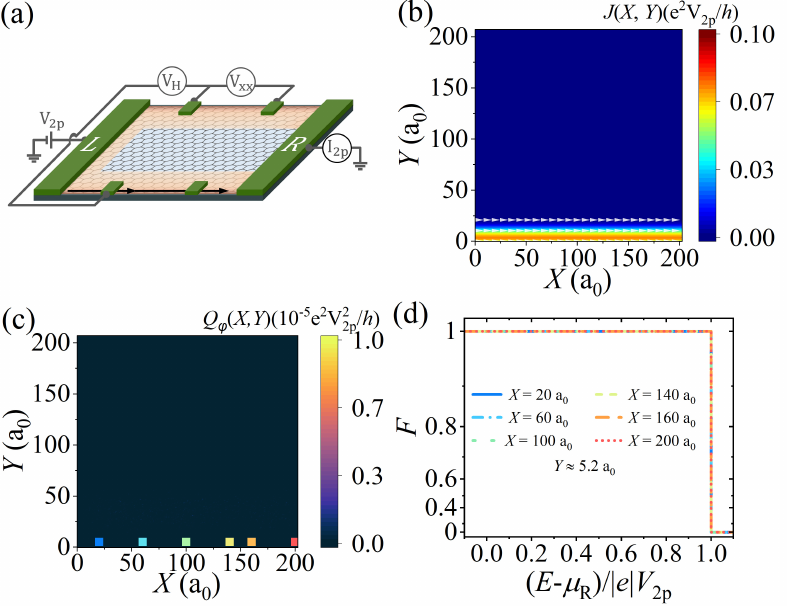}
 \caption{(a). Schematic plot of a six-terminal Hall device in graphene.
 In contrast to Fig.~2 in the main text, the absence of the elastic barrier $\mathcal{V}_n$ is taken into account.
 An edge electrostatic potential $\mathcal{E}^{\mathrm{edge}}_{n}$ is applied in the orange region, which simulates the edge reconstruction in realistic materials.
 The chemical potential of contact $\mathrm{L}(\mathrm{R})$ is represented by $\mu_{\mathrm{L}}(\mu_{\mathrm{R}})$.
 (b) The local electric current $J(X,Y)$ and (c) the local energy current $Q_{\varphi}(X,Y)$ in the quantum Hall plateau regime with $R_{xy}=-h/e^2$.
 Here, $E_\mathrm{F}=0.05~\mathrm{eV}$ and $\Phi_{\mathrm{B}}=0.004$.
 The B{$\mathrm{\ddot{u}}$}ttiker probes are randomly distributed all over the graphene sample and the numerical simulation is averaged over 1280 B{$\mathrm{\ddot{u}}$}ttiker-probe configurations. 
 The energy distribution of electrons $F$ as a function of normalized energy $(E-\mu_\mathrm{R})/\abs{e}V_\mathrm{2p}$ (0 for the right contact and 1 for the left contact) in (d) refers to the sites marked by colored squares in (b).}
 \label{noscattering}
\end{figure}

In the main text, we introduce an elastic barrier denoted as $\mathcal{V}_n$, which is positioned at the lower edge of graphene. The scattering barrier takes the form of a Gaussian-shaped long-range potential, with its mathematical expression as follows:
\begin{equation}
 \mathcal{V}_n =\mathcal{V}_0e^{-\abs{\mathbf{R}_n-\mathbf{R}_0}/\Lambda}
\end{equation}
where $\mathbf{R}_n=(x_n,~y_n)$ labels the coordinate of lattice site $n$, $\mathbf{R}_0$ denotes the potential center and $\mathcal{V}_0$ describes the scattering strength. In calculations, $\mathbf{R}_0\approx(105a_0, 29.7a_0)$, $\mathcal{V}_0=-0.35\mathrm{eV}$ and $\Lambda=10a_0$.
Here, in~\cref{noscattering}, the absence of the elastic barrier $\mathcal{V}_n$ is taken into account.
Then, the electric current directly flows from the left contact into the right contact [see the local current density $J(X, Y)$ in~\cref{noscattering}(b)].
Following the semiclassical analysis in the main text, electrons occupying the outer channel stay in equilibrium with the left contact, i.e., a zero-temperature Fermi distribution with chemical potential~$\mu_{\mathrm{L}}$.
As in~\cref{noscattering}(d), the energy distribution of electrons $F(n, E)$ is exactly a step function $F(E) = \theta(\mu_{\mathrm{L}}-E)$.
Even though the energy relaxation processes are included, the electrons in equilibrium do not dissipate their energy into environmental reservoirs, which is consistent with the numerical result of local energy current $Q_{\varphi}(X, Y)$ in~\cref{noscattering}(c).

In this context, energy dissipation does not occur because non-equilibrium electrons are absent.
We show that the elastic barrier in the main text causes mixing between the inner and outer edge channels, which enables the production of non-equilibrium electrons that are crucial for heat generation.
In real devices, elastic scattering processes simulated by the barrier $\mathcal{V}_n$ are ubiquitous.
% The absence of heat dissipation confirms that the energy dissipation is indeed sourced from the edge states.
\clearpage
\pagebreak
\section{A case with the elastic barrier $\mathcal{V}_n$ in the bulk}
\begin{figure}[htbp]
 \includegraphics[width=.6\columnwidth]{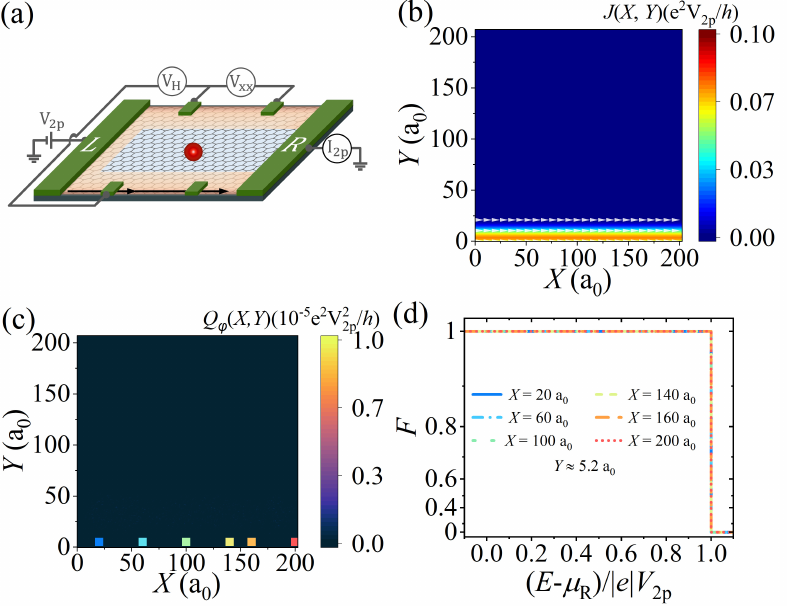}
 \caption{(a). Schematic plot of a six-terminal Hall device in graphene.
 In contrast to Fig.~2 in the main text, the elastic barrier $\mathcal{V}_n$ is placed within the bulk.
 An edge electrostatic potential $\mathcal{E}^{\mathrm{edge}}_{n}$ is applied in the orange region, which simulates the edge reconstruction in realistic materials.
 The chemical potential of contact $\mathrm{L}(\mathrm{R})$ is represented by $\mu_{\mathrm{L}}(\mu_{\mathrm{R}})$.
 (b) The local electric current $J(X,Y)$ and (c) the local energy current $Q_{\varphi}(X,Y)$ in the quantum Hall plateau regime with $R_{xy}=-h/e^2$.
 Here, $E_\mathrm{F}=0.05~\mathrm{eV}$ and $\Phi_{\mathrm{B}}=0.004$.
 The B{$\mathrm{\ddot{u}}$}ttiker probes are randomly distributed all over the graphene sample and the numerical simulation is averaged over 1280 B{$\mathrm{\ddot{u}}$}ttiker-probe configurations. 
 The energy distribution of electrons $F$ as a function of normalized energy $(E-\mu_\mathrm{R})/\abs{e}V_\mathrm{2p}$ (0 for the right contact and 1 for the left contact) in (d) refers to the sites marked by colored squares in (b).}
 \label{bulkscattering}
 \end{figure}

Compared with the previous section and the main text, we introduce an elastic barrier denoted as $\mathcal{V}_n$, which is positioned in the bulk of graphene. In calculations, the potential center $\mathbf{R}_0\approx(105a_0, 110a_0)$ and $a_0$ is the lattice constant of graphene.
The electric current directly flows from the left contact into the right contact [see the local current density $J(X, Y)$ in~\cref{bulkscattering}(b)]. Following the semiclassical analysis in the main text, electrons occupying the outer channel stay in equilibrium with the left contact, i.e., a zero-temperature Fermi distribution with chemical potential~$\mu_{\mathrm{L}}$. As in~\cref{bulkscattering}(d), the energy distribution of electrons $F(n, E)$ is exactly a step function $F(E) = \theta(\mu_{\mathrm{L}}-E)$. Even though the energy relaxation processes are included, the electrons in equilibrium do not dissipate their energy into environmental reservoirs, which is consistent with the numerical result of local energy current $Q_{\varphi}(X, Y)$ in~\cref{bulkscattering}(c).
The absence of heat dissipation confirms that the energy dissipation is indeed sourced from the edge states.
\clearpage
{\section{Dissipation analysis of a Chern junction}}
\begin{figure}[htbp]
 \includegraphics[width=0.99\columnwidth]{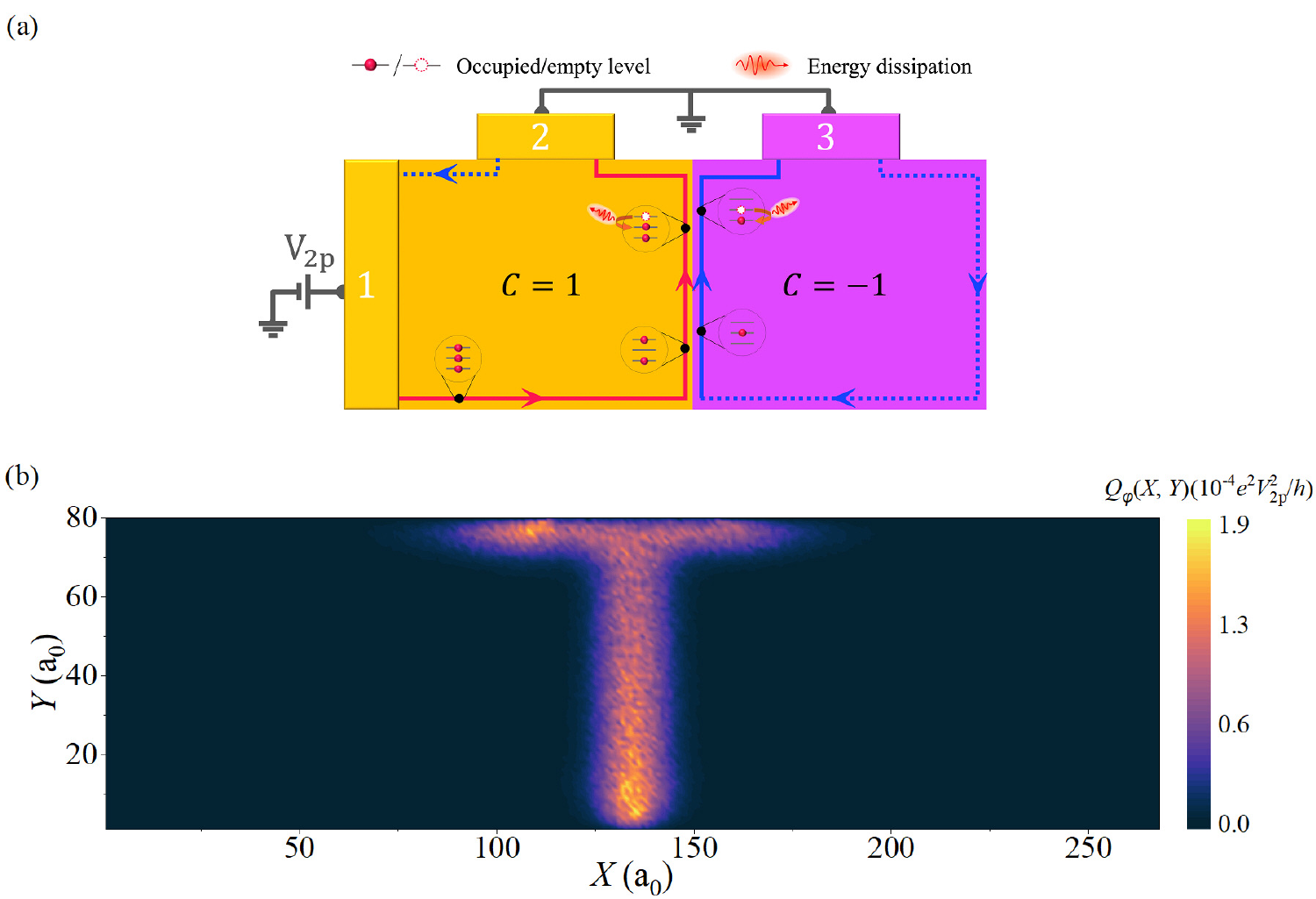}
 \caption{(a) Schematic plot of the chiral current divider reproduced from the Fig.~3b of the reference (Nat. Commun. \textbf{14}, 770 (2023)~\cite{zhaocreationchiralinterface2023}). (b) The calculated result of local energy dissipation which is averaged over 216 B{$\mathrm{\ddot{u}}$}ttiker-probe configurations.}\label{fig:divider}
\end{figure}
Regarding two recent experiments [Nat. Commun. \textbf{13}, 5967 (2022); Nat. Commun. \textbf{14}, 770 (2023)], they have reported the realization of quantum anomalous Hall (QAH) junctions with different Chern numbers and observed quantized charge transport. Upon these findings, they claim that junctions composed of Chern insulators can operate as dissipation-free current dividers, and hold potential dissipationless topological devices. 

In the last paragraph of the section ``\textit{Experimental routines and discussions}'' in the main text, we predict that energy dissipation without backscattering can still emerge inside the conductors in these experimental devices [e.g., Nat. Commun. \textbf{13}, 5967 (2022); Nat. Commun. \textbf{14}, 770 (2023)]. To make such a statement more solid, we apply our theory to one of the devices [see \cref{fig:divider}(a)] which is a junction consisting of Chern insulators. In this device, terminal 2 and 3 are grounded and a chemical potential difference is applied between terminal 1 and them. The two-resistance (from terminal 1 to ground) is also quantized ($R_{2p}=h/e^2$), i.e., there is no backscattering and chiral edge transport eliminates local resistivity leaving only quantized contact resistance. However, significant energy dissipation emerges at the domain wall of the junction as well as its downstream region [see \cref{fig:divider}(b)].

An intuitive and semiclassical explanation is consistent with the physical picture in our manuscript [see the schematic of the transport process and qualitative energy distributions of electrons in \cref{fig:divider}(a)]. The red channel to the left of the domain wall is occupied only by electrons coming in from the left contact. Hence these states are in equilibrium and have the same electrochemical potential as the left contact [see the schematic of fully occupied energy levels at the bottom-left of \cref{fig:divider}(a)]. At the domain wall, electrons are scattered between two chiral edge channels. Hence, in the energy range between $\mu_{low}$ and $\mu_{high}$, the states are filled partially [see the schematic of partially occupied levels at the bottom-middle of \cref{fig:divider}(a)]. Such a distribution is a highly nonequilibrium one. Through inelastic processes such as phonon emission, the electrons will settle down to lower energies and a Fermi distribution will be established [see the schematic of occupation in new equilibrium at the top-middle of \cref{fig:divider}(a)]. This evolution process leads to the dissipation of energy. Thus, the energy dissipation in the quantum limit is universal in topological devices and is especially important for the application of topological materials in dissipationless electronic devices.
\clearpage
\pagebreak
\section{Recursive Green's function method}

The primary physical quantities can be represented in terms of Green's
functions. Generally speaking, to obtain Green's functions, it is
necessary to compute the inverse of large matrices, which is usually
infeasible in practice. However, the recursive Green's function method
is an efficient algorithm for calculating Green's functions~\cite{lewenkopfRecursiveGreenFunction2013,cattenaGeneralizedMultiterminalDecoherent2014}.

For a tight-binding model, the effective Hamiltonian, $\widehat{H}_{\mathrm{eff}}$,
that includes real leads and B$\mathrm{\ddot{u}}$ttiker probes, reads
\[
\widehat{H}_{\mathrm{eff}}=\widehat{H}+\widehat{\Sigma}_{\phi},
\]
where $\widehat{H}$ refers to the Hamiltonian consisting of the contacts,
the central region and the hopping matrix between contacts and the
central region. $\widehat{\Sigma}_{\phi}$, as in the main text, represents
the self-energies accounting for the inclusion of B$\mathrm{\ddot{u}}$ttiker
probes.

By partitioning the device into $N$ slices and doing mathematical
derivation, one can obtain the following specific form~\cite{cattenaGeneralizedMultiterminalDecoherent2014}.
For the diagonal blocks, we have 
\begin{subequations}
\begin{eqnarray}
G_{ii} & = & \left[\left(\epsilon\mathbf{I}-\mathbf{H}_{i}\right)-\Sigma_{i}^{(1)}-\Sigma_{i}^{(N)}\right]^{-1},\label{eq:diagonal_G_ii}\\
G_{jj} & = & \left[\left(\epsilon\mathbf{I}-\mathbf{H}_{j}\right)-\Sigma_{j}^{(1)}-\Sigma_{j}^{(N)}\right]^{-1},\label{eq:diagonal_G_jj}
\end{eqnarray}
\end{subequations}
 where $\mathbf{H}_{i}$ refers to the Hamiltonian of the $i$-th
slice when it is isolated from the system. Additionally, $\Sigma_{i}^{(j)},\Sigma_{j}^{(i)}$
have the forms 
\begin{subequations}
\begin{eqnarray}
\Sigma_{i}^{(j)} & = & \left[\mathbf{V}_{i,i+1}\left(\epsilon\mathbf{I}-\mathbf{H}_{i+1}-\Sigma_{i+1}^{(j)}\right)^{-1}\right]\mathbf{V}_{i+1,i},\label{eq:sigma_1}\\
\Sigma_{j}^{(i)} & = & \left[\mathbf{V}_{j,j-1}\left(\epsilon\mathbf{I}-\mathbf{H}_{j-1}-\Sigma_{j-1}^{(i)}\right)^{-1}\right]\mathbf{V}_{j-1,j},\label{eq:sigma_2}\\
 & \mathrm{for} & j>i\nonumber 
\end{eqnarray}
\end{subequations}
\\
where $\mathbf{V}_{i,i+1}$ is the hopping matrix from the $i$-th
slice to the $i\!+\!1$-th slice. The off-diagonal block of the Green's
function reads\\
\begin{subequations}
\begin{eqnarray}
G_{ij} & = & G_{ii}\prod_{k=i}^{j-1}\left[\Sigma_{k}^{(N)}\mathbf{V}_{k+1,k}^{-1}\right],\label{eq:off_diagonal_G_ij}\\
G_{ji} & = & G_{jj}\prod_{k=j}^{i+1}\left[\Sigma_{k}^{(1)}\mathbf{V}_{k-1,k}^{-1}\right].\label{eq:off_diagonal_G_ji}\\
 & \mathrm{for} & j>i\nonumber 
\end{eqnarray}
\end{subequations}
Thus, equations \eqref{eq:diagonal_G_ii}-\eqref{eq:off_diagonal_G_ji}
constitute the kernel of the recursive Green's function method.

\clearpage
\pagebreak
\section{Charge transport}

\subsection{Landauer formula of charge current}
First, we consider a multi-terminal device.
In this case, the charge current flowing out of terminal $p$ can be defined as the rate of change of particle number~\cite{jauhoTimedependentTransportInteracting1994},

\begin{eqnarray*}
J_{p} & = & -e\left\langle \dot{N}_{p}\right\rangle \\
 & = & -e\left\langle \frac{\mathrm{d}N_{p}}{\mathrm{d}t}\right\rangle \\
 & = & \frac{ie}{\hbar}\left\langle \left[N_{p},H\right]\right\rangle 
\end{eqnarray*}
where $N_{p}=\sum_{k,\alpha\in p}C_{k\alpha}^{\dagger}C_{k\alpha}$.
After some straightforward calculations, one can readily find

\begin{equation*}
\left[N_{p},H\right]=\sum_{k,\alpha\in p,\,n}\left(\mathbf{V}_{k\alpha,n}\left\langle C_{k\alpha}^{\dagger}d_{n}\right\rangle -\mathbf{V}_{n,k\alpha}\left\langle d_{n}^{\dagger}C_{k\alpha}\right\rangle \right).
\end{equation*}
where $\mathbf{V}_{k\alpha,n}$ is the hopping matrix between the state $\left(k,\alpha\right)$in contact $p$ and the state $n$ in the central region.
Thus, 

\begin{equation*}
J_{p}=\frac{2e}{\hbar}\Re\left[\sum_{k,\alpha\in p,\,n}\mathbf{V}_{k\alpha,n}G_{n,k\alpha}^{<}\left(t,t\right)\right].
\end{equation*}
with the definition $G_{n,k\alpha}^{<}\left(t,t'\right)\equiv i\left\langle C_{k\alpha}^{\dagger}\left(t'\right)d_{n}\left(t\right)\right\rangle $.
$\Re\left[\dots\right]$ means taking the real part.

By utilizing the equation-of-motion technique, the Fourier transformation
of $G_{n,k\alpha}^{<}\left(t,t'\right)$ is

\begin{equation*}
G_{n,k\alpha}^{<}\left(\epsilon\right)=\sum_{m}\mathbf{V}_{k\alpha,m}^{*}\left[G_{nm}^{r}\left(\epsilon\right)g_{k\alpha}^{<}\left(\epsilon\right)+G_{nm}^{<}\left(\epsilon\right)g_{k\alpha}^{\alpha}\left(\epsilon\right)\right].
\end{equation*}
So, the current becomes

\begin{equation*}
J_{p}=\frac{2e}{\hbar}\Re\int\frac{\mathrm{d}\epsilon}{2\pi}\sum_{k,\alpha\in p,\,n,m}\mathbf{V}_{k\alpha,n}\mathbf{V}_{k\alpha,m}^{*}\left[G_{nm}^{r}\left(\epsilon\right)g_{k\alpha}^{<}\left(\epsilon\right)+G_{nm}^{<}\left(\epsilon\right)g_{k\alpha}^{a}\left(\epsilon\right)\right].
\end{equation*}
Converting the momentum summation to energy integration, it is easy
to find

\begin{eqnarray*}
 & & \sum_{k,\alpha\in p}\mathbf{V}_{k\alpha,n}\mathbf{V}_{k\alpha,m}^{*}g^{<}\left(\epsilon\right)\\
 & = & \sum_{k,\alpha\in p}\mathbf{V}_{k\alpha,n}\mathbf{V}_{k\alpha,m}^{*}if_{p}\left(\epsilon_{k\alpha}\right)2\pi\delta\left(\epsilon-\epsilon_{k\alpha}\right)\\
 & = & if_{p}\left(\epsilon\right)\Gamma_{mn}^{p}\left(\epsilon\right)
\end{eqnarray*}
with the definition of a line-width function $\Gamma_{mn}^{p}\left(\epsilon\right)=2\pi\sum_{\alpha}\rho_{\alpha}\left(\epsilon\right)\mathbf{V}_{k\alpha,n}\mathbf{V}_{k\alpha,m}^{*}$
and the Fermi distribution function $f_{p}\left(\epsilon\right)$.
For the other term,

\begin{eqnarray*}
 & & \sum_{k,\alpha\in p}\mathbf{V}_{k\alpha,n}\mathbf{V}_{k\alpha,m}^{*}g_{k\alpha}^{a}\left(\epsilon\right)\\
 & = & \sum_{k,\alpha\in p}\mathbf{V}_{k\alpha,n}\mathbf{V}_{k\alpha,m}^{*}/\left(\epsilon-\epsilon_{k\alpha}-i0^{+}\right)\\
 & = & \sum_{k,\alpha\in p}\mathbf{V}_{k\alpha,n}\mathbf{V}_{k\alpha,m}^{*}i\pi\delta\left(\epsilon-\epsilon_{k\alpha}\right)\\
 & = & \frac{i}{2}\Gamma_{mn}^{p}\left(\epsilon\right)
\end{eqnarray*}
So, the current becomes

\begin{eqnarray}
J_{p} & = & \frac{2e}{\hbar}\Re\int\frac{\mathrm{d}\epsilon}{2\pi}\sum_{n,m}\left[G_{nm}^{r}\left(\epsilon\right)if_{p}\left(\epsilon\right)\Gamma_{mn}^{p}\left(\epsilon\right)+\frac{i}{2}G_{nm}^{<}\left(\epsilon\right)\Gamma_{mn}^{p}\left(\epsilon\right)\right]\nonumber \\
 & = & -\frac{2e}{\hbar}\Im\int\frac{\mathrm{d}\epsilon}{2\pi}\mathrm{Tr}\left\{ \Gamma^{p}\left(\epsilon\right)\left[f_{p}\left(\epsilon\right)G^{r}\left(\epsilon\right)+\frac{1}{2}G^{<}\left(\epsilon\right)\right]\right\} \label{eq:current_Gless}
\end{eqnarray}
where $\Im\left[\dots\right]$ means taking the imaginary part. According to the Keldysh's
equation and the fluctuation-dissipation theorem, the second term
in \eqref{eq:current_Gless} is

\begin{eqnarray}
G^{<} & = & G^{r}\Sigma^{<}G^{a}\nonumber \\
 & = & G^{r}\left(\sum_{q}\Sigma_{q}^{<}\right)G^{a}\nonumber \\
 & = & -\sum_{q}G^{r}f_{q}\left(\Sigma_{q}^{r}-\Sigma_{q}^{a}\right)G^{a}\nonumber \\
 & = & \sum_{q}G^{r}if_{q}\Gamma^{q}G^{a}.\label{eq:gless}
\end{eqnarray}
For the first term,

\begin{eqnarray*}
2\Im\mathrm{Tr}\left(\Gamma^{p}G^{r}\right) & = & -i\mathrm{Tr}\left[\Gamma^{p}\left(G^{r}-G^{a}\right)\right]\\
 & = & -i\mathrm{Tr}\left[\Gamma^{p}G^{r}\left(\Sigma^{r}-\Sigma^{a}\right)G^{a}\right]\\
 & = & -\mathrm{Tr}\left[\sum_{q}\Gamma^{p}G^{r}\Gamma^{q}G^{a}\right]
\end{eqnarray*}
Finally, the current flowing out of contact $p$ is

\begin{eqnarray*}
J_{p} & = & \frac{e}{\hbar}\int\frac{\mathrm{d}\epsilon}{2\pi}\sum_{q}\left[f_{p}\left(\epsilon\right)-f_{q}\left(\epsilon\right)\right]\mathrm{Tr}\left(\Gamma^{p}G^{r}\Gamma^{q}G^{a}\right)\nonumber \\
 & = & \frac{e}{h}\sum_{q}\int T_{pq}\left(\epsilon\right)\left[f_{p}\left(\epsilon\right)-f_{q}\left(\epsilon\right)\right]\mathrm{d}\epsilon\label{eq:final_J}
\end{eqnarray*}
where $T_{pq}\left(\epsilon\right)\equiv\mathrm{Tr}\left(\Gamma^{p}G^{r}\Gamma^{q}G^{a}\right)$
is the transmission matrix. Under low bias and low temperature, by
Taylor's expansion,

\begin{eqnarray*}
J_{p} & = & \frac{e^{2}}{\hbar}\sum_{q}T_{pq}\left(\epsilon_{\mathrm{F}}\right)\left(V_{p}-V_{q}\right)\mathrm{d}\epsilon
\end{eqnarray*}
which is the Eq. (2) in the main text.

\subsection{Local electric current}

For a two-terminal device, i.e., only with source and drain contacts,
the local current flowing from site $\mathbf{i}$ can be expressed
as~\cite{jiangNumericalStudyTopological2009}

\begin{eqnarray*}
J_{i} & = & -e\left\langle \dot{N}_{i}\right\rangle \\
 & = & -e\left\langle \frac{\mathrm{d}N_{i}}{\mathrm{d}t}\right\rangle \\
 & = & \frac{ie}{\hbar}\left\langle \left[\sum_{\alpha}C_{i\alpha}^{\dagger}C_{i\alpha},H\right]\right\rangle \\
 & = & \frac{2e}{\hbar}\sum_{j}\sum_{\alpha,\beta}\Re\left[\mathbf{V}_{i\alpha,j\beta}G_{j\beta,i\alpha}^{<}\left(t,t\right)\right]
\end{eqnarray*}
where $\alpha,\beta$ label the internal degrees of freedom. By similar
calculation, the local current from site $\mathbf{i}$ to site $\mathbf{j}$
reads 

\begin{eqnarray*}
J_{i\rightarrow j} & = & \frac{2e}{\hbar}\Re\int\frac{\mathrm{d}\epsilon}{2\pi}\sum_{\alpha,\beta}\left[\mathbf{V}_{i\alpha,j\beta}G_{j\beta,i\alpha}^{<}\left(\epsilon\right)\right]
\end{eqnarray*}
Substitute \eqref{eq:gless} into it, one can obtain

\begin{eqnarray}
J_{i\rightarrow j} & = & -\frac{2e}{h}\sum_{\alpha,\beta}\int_{-\infty}^{+\infty}\mathrm{d}\epsilon\Im\left\{ \mathbf{V}_{i\alpha,j\beta}\left[G^{r}\left(f_{\mathrm{L}}\Gamma^{\mathrm{L}}+f_{\mathrm{R}}\Gamma^{\mathrm{R}}\right)G^{a}\right]_{j\beta,i\alpha}\right\} \nonumber \\
 & = & -\frac{2e}{h}\sum_{\alpha,\beta}\int_{-\infty}^{eV_{\mathrm{R}}}\mathrm{d}\epsilon\Im\left\{ \mathbf{V}_{i\alpha,j\beta}\left[G^{r}\left(f_{\mathrm{L}}\Gamma^{\mathrm{L}}+f_{\mathrm{R}}\Gamma^{\mathrm{R}}\right)G^{a}\right]_{j\beta,i\alpha}\right\} \nonumber \\
 & & -\frac{2e^{2}}{h}\sum_{\alpha,\beta}\Im\left\{ \mathbf{V}_{i\alpha,j\beta}\left[G^{r}\Gamma^{\mathrm{L}}G^{a}\right]_{j\beta,i\alpha}\right\} \left(V_{\mathrm{L}}-V_{\mathrm{R}}\right)\label{eq:local_current}
\end{eqnarray}
The last line is evaluated at zero temperature.

The local electric current $J\left(X,Y\right)$ in Fig.2(a) of the main text refers to the current flowing along the carbon-carbon bond located at $\left(X,Y\right)$. In a graphene device, each carbon-carbon bond connects a pair of carbon atoms $\mathbf{i}$ and $\mathbf{j}$.
So, $J\left(X,Y\right)$ is actually the quantity $J_{i\rightarrow j}$ calculated by \eqref{eq:local_current}.

\clearpage
\pagebreak
\section{Energy current transport}

\subsection{Energy current}

Firstly, we can define the energy flow. Intuitively,
electric current can be defined as 

\begin{eqnarray*}
J & = & e\sum_{\alpha}n_{\alpha}v_{\alpha}
\end{eqnarray*}
where $\alpha$ labels the quantum states. Electric current
represents the number of electrons passing through a unit area in
a unit of time. Each particle carries a charge $e$. Thus, the energy
current density can be similarly defined as

\begin{eqnarray*}
J_{E} & = & \sum_{\alpha}\epsilon_{\alpha}n_{\alpha}v_{\alpha}
\end{eqnarray*}
where charge $e$ is replaced with the energy $\epsilon_{\alpha}$
carried by the particle. The multi-terminal Landauer formula of energy
current flowing into terminal $p$ reads

\begin{eqnarray*}
J_{E_{p}} & = & \frac{e}{h}\sum_{q}\int\epsilon T_{pq}\left(\epsilon\right)\left[f_{p}\left(\epsilon\right)-f_{q}\left(\epsilon\right)\right]\mathrm{d}\epsilon
\end{eqnarray*}
\subsection{Local energy dissipation}
Under the B$\mathrm{\ddot{u}}$ttiker-probe scheme, local energy dissipation is defined as the difference between the energy
current flowing into terminal $p$, $J_{E_{p}}$, and the energy current
in the environmental reservoir, $\mu_{p}J_{p}$,

\begin{eqnarray*}
Q_{p} & = & \mu_{p}J_{p}-J_{E_{p}}\\
 & = & -\frac{1}{h}\sum_{q}\int\left(\epsilon-\mu_{p}\right)T_{pq}\left(\epsilon\right)\left[f_{p}\left(\epsilon\right)-f_{q}\left(\epsilon\right)\right]\mathrm{d}\epsilon
\end{eqnarray*}

Under low bias and low temperature, one can obtain the Eq.(2) in the
main text,

\begin{eqnarray*}
Q_{p} & = & \frac{e^{2}}{2h}\sum_{q}T_{pq}(V_{p}-V_{q})^{2}-\frac{\pi^{2}k_{\mathrm{B}}^{2}}{6h}\sum_{q}T_{pq}(T_{p}^{2}-T_{q}^{2}).
\end{eqnarray*}

The local energy dissipation $Q\left(X,Y\right)$ in the Fig. 2(b) of the main text refers to the net energy current flowing into the B{$\mathrm{\ddot{u}}$}ttiker probe $p$ which is located at $\left(X,Y\right)$. The Fig. 2(b) in the main text shows the averaged $Q\left(X,Y\right)$ over different B{$\mathrm{\ddot{u}}$}ttiker-probe configurations.

\section{Energy distribution of electrons}

With Green's functions, the energy distribution of electrons can be
defined intuitively as~\cite{fangThermalDissipationQuantum2021},

\begin{eqnarray}
F(n,E) & = & \frac{\mathrm{Tr\left[G_{n,n}^{e}\left(E\right)\right]}}{\mathrm{Tr\left[A_{n,n}\left(E\right)\right]}}.
\end{eqnarray}
The spectral function is $A_{n,n}\left(E\right)=i\left[G_{n,n}^{r}\left(E\right)-G_{n,n}^{a}\left(E\right)\right]$,
which tells us the nature of the allowed electronic states, regardless
of whether they are occupied or not. $G_{n,n}^{e}\left(E\right)=-iG_{n,n}^{<}\left(E\right)$
is the electron correlation function which tells how many of these
states are occupied or empty. Thus, the ratio of them reflects the
distribution probability. Here,

\begin{subequations}
 \begin{eqnarray*}
 G^{<} & = & G^{r}\Sigma^{<}G^{a}\\
 \Sigma^{<} & = & \sum_{q}if_{q}\Gamma^{q}\\
 G^{a} & = & \left(G^{r}\right)^{\dagger}
 \end{eqnarray*}
\end{subequations}
where $G^{r}$ can be calculated numerically by recursive Green's
function method.

\clearpage
\pagebreak
\section{Local density of states}
\subsection{Theoretical formula}
In terms of Green's function, the local density of states $\rho$ at a given position $r$ and a given Fermi energy $E_\mathrm{F}$ is defined as follows:

\begin{equation}
\rho\left(r,E\right) = \frac{1}{2\pi}\mathrm{tr}\left[A\left(r,E\right)\right]\label{eq:ldos}
\end{equation}
where the spectral function is $A\left(E\right)=i\left[G^{r}\left(E\right)-G^{a}\left(E\right)\right]$. The diagonal elements of the spectral function give the local density of states.
\subsection{Relevant analysis about our model}
\begin{figure}[htbp]
 \includegraphics[width=0.99\columnwidth]{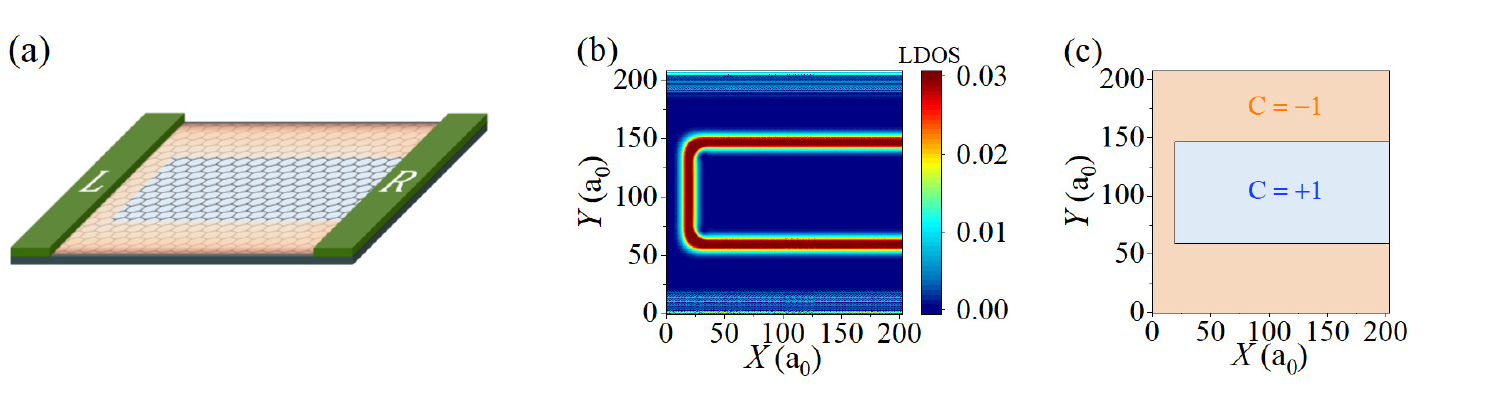}
 \caption{(a) Schematic plot of the graphene device attached with semi-infinite left and right contacts. An edge electrostatic potential $\mathcal{E}^{\mathrm{edge}}_{n}=0.1~\mathrm{eV}$, as in the main text, is applied in the orange region with $E_{\mathrm{F}}=0.05~\mathrm{eV}$ and $\Phi_{\mathrm{B}}=0.004$. (b) Local density of states (\textrm{LDOS}) of the central region without B$\mathrm{\ddot{u}}$ttiker probes and the scattering barrier. (c) Chern numbers of different regions in (a).}\label{fig:ldos}
\end{figure}
we consider the device as depicted in Figure 2(a) of the main text, without the inclusion of side electrodes, as shown in \cref{fig:ldos}(a). Within our physical framework, in contrast to perfect graphene, an electrostatic potential $\mathcal{E}^{\mathrm{edge}}_{n}$ is added in the vicinity of the graphene boundary [see the orange region in \cref{fig:ldos}(a)]. To better illustrate the structure of edge states, the long-range scattering potential barrier has been removed. Under a magnetic field, Landau levels will form in the device. By recursive Green's function method, the local density of states at a given position and a given Fermi energy can be calculated by \cref{eq:ldos}.

The calculation results are presented in \cref{fig:ldos}(b), where the edge potential results in the existence of edge states on both the inner and outer sides of the sample, i.e., the inner and outer channels. If there were no edge potential barrier, the system would be in a phase with a Chern number of 1. However, due to the presence of the edge potential barrier, the energy in the orange region is shifted, leading to a phase with a Chern number of -1. According to the bulk-edge correspondence, the environment outside the device is equivalent to vacuum, resulting in one edge state along the outer boundary of the device. Additionally, two edge states are found at the domain wall within the device's interior.
% \bibliography{reference.bib}